\documentclass[aps,twocolumn,pra,superscriptaddress,amsmath,showpacs]{revtex4}
\usepackage{slashed,bbm,hyperref,amsmath,amssymb,amsfonts,graphicx,array}
\graphicspath{{Figures/}}
\usepackage{float,subfig,xcolor}
\usepackage[english]{babel}
\allowdisplaybreaks[1-4]
\usepackage{bigstrut}
\usepackage{multirow}
\newtheorem{theorem}{Theorem}
\newcommand{\sig}{\sigma}
\newcommand{\calh}{\mathcal{H}}
\newcommand{\beq}{\begin{equation}}
\newcommand{\eeq}{\end{equation}}
\newcommand{\bea}{\begin{eqnarray}}
\newcommand{\eea}{\end{eqnarray}}
\newcommand{\til}{\tilde}
\newcommand{\rhot}{\tilde\rho}
\newcommand{\rg}{\rangle}
\newcommand{\dg}{\dagger}

\newcommand{\bsplit}{\begin{split}}
\begin{document}

\title{Nonequilibrium Effects on Quantum Correlations: Discord, Mutual Information and Entanglement of a Two-Fermionic System in Bosonic and Fermionic Environments}

\author{Xuanhua Wang}
\email[]{xuanhua.wang@stonybrook.edu}
\affiliation{Department of Physics and Astronomy, SUNY, Stony Brook}
\author{Jin Wang}
\email[]{Corresponding Author: jin.wang1@stonybrook.edu}
\affiliation{Department of Physics and Astronomy, SUNY, Stony Brook}
\affiliation{Department of Chemistry, SUNY, Stony Brook }

\date{\today}
\begin{abstract}
We study the steady state entanglement and correlations of an open system comprised of two coupled fermions in the equilibrium or nonequilibrium environments and distill the nonequilibrium contribution to the quantum correlations. We show that in the equilibrium condition, the steady-state quantum correlations exhibit non-monotonic behavior, while in the nonequilibrium case, the monotonicity is determined by many parameters. The entanglement vanishes abruptly upon the increase of the temperature (bias) and chemical potential bias, it witnesses a critical chemical potential above which the concurrence always remains positive. In the fermionic reservoirs, quantum correlations reach the maximal values when one chemical potential of the reservoirs matches to the system frequency. We separate the quantum correlation generation due to the averaged effect from the pure nonequilibrium effect. In contrast with the previous results, when the averaged effect is separated out, the nonequilibrium generation of quantum correlation shows a distinctive monotonic behavior. The difference between the large-tunneling regime with decaying correlations and the small-tunneling with increasing correlations is discussed. Near the boundary of the two regimes, the entanglement behavior is a mixture of two extremes, it resurrects with the increase of chemical potential bias after its previous drop to zero. 
\end{abstract}

\pacs{03.65.Ud, 03.65.Yz, 03.67.-a, 05.70.Ln}

\maketitle

\numberwithin{equation}{section}
\section{Introduction}
Entanglement is one of the most widely-used measures of nonclassicality in a quantum system. Nevertheless, in many cases unentangled states can also exhibit nonclassical behavior \cite{boundary,review2,zurek}. In particular, when dealing with \textit{mixed states}, the definition of entanglement can be generalized to the weighted sum of the pure state entanglement in the decomposition of the mixed states minimized over all decomposition. This minimization forces the entanglement to vanish when a certain disorder of the system is attained \cite{compare, cp1,cp2,cp3}. Thus augmenting measures of quantum correlations other than entanglement provides a more complete picture to capture the difference between the quantum and classical worlds \cite{boundary, zurek}.

Quantum discord (QD) is used as a characterization of quantum correlation in information theory, quantum computing and also biophysics due to its robustness with respect to the noises from the environments. Although its merit in characterizing the speedup of deterministic quantum computation of one qubit was disputed \cite{knill, question0, question}, it an important quantity for quantifying correlations beyond the classical connections. Quantum discord has fertile applications in studying biological systems as photosynthesis in the light-harvesting pigment-protein complexes and tunnelling through enzyme-catalysed reactions \cite{review1, 371,372,373,374}. Despite its application, computing quantum discord is a NP-complete problem \cite{NP}. Even for a general two-qubit state, getting a closed analytical form of discord is still challenging. On the other hand, the concurrence used for describing a four dimensional density matrix is relatively simple and unambiguous. Many previous researches have explored the properties of quantum discord with certain designated density matrices \cite{ali, luo, Fei}. The discord in special cases such as a few-parameter families \cite{luo, Sarandy, RD, errorqd} and in “X" form of density matrix in higher dimensions \cite{highd} have been calculated. Limited studies have devoted to open systems\cite{lgi}. Distinct behaviors of entanglement and discord were noticed for a simple two qubit system under environmental biases \cite{lgi, segal}. 

Dissipative quantum systems exhibited many non-smooth behaviors such as the long-lived quantum coherence  \cite{zd1,sl1}, the \textit{sudden death} of entanglement \cite{YuEberly2,review1, YuEberly} and \textit{transition} from quantum to classical world through decoherence measured by discord \cite{transition}. Similar abrupt behaviors as in the relaxation process were also witnessed in the steady states and also in the nonequilibrium conditions with the continuously varying environmental parameters \cite{ali,lgi,segal}. It was shown that while entanglement manifests a similar "sudden death", the quantum discord vanishes only in the asymptotic limit \cite{suddendeath, segal, lgi}. In this paper, we show that in some particular parameter regimes, entanglement can \textit{resurrect} after its previous drop to zero as we tune up the chemical potential. Besides, the quantum correlations do not necessarily decay to zero as the bias of the environments increases but can have large asymptotic value in some parameter regimes.

\textit{Nonequilibrium} effects for the quantum correlations in an open system was also discussed by previous studies \cite{zd2,sl2,lgi,wuwei1,thermal}. Some of the previous results showed that the bias of the reservoirs can only deteriorate quantum correlations in the strong-coupling model \cite{segal} or can minutely increase the correlations in weak-coupling model \cite{lgi,wuwei1,thermal}. However, whether the effect is truly due to the nonequilibriumness or rather an averaged effect of different reservoirs was not differentiated. In fact, we find that the nonequilibrium effects discussed in the previous studies \cite{segal,lgi,wuwei1,thermal} are mainly due to the system being in equilibrium with the \textit{sum} or \textit{average} of different reservoirs, while the most intrinsic nature of the nonequilibrium effect which should come from the \textit{difference} between two reservoirs was not captured. In this work, we see that the environments do not act as a peripheral role, but can greatly change the trend and strength of the quantum correlations. We separate the component of the quantum correlation generation due to the essential nonequilibriumness from that due to the averaged effect of the environments. The \textit{``distilled" nonequilibrium contribution} after separation has a distinctive feature that monotonically boosts the quantum correlations and entanglement. Our result suggests that the nonequilibrium effect, which is left out from the Lindblad formalism through ignoring the coherence terms \cite{wuwei1,lind}, is important in studying the quantumness and correlations of quantum systems.

In general, the relationships among quantum discord, entanglement, and classical correlation are not yet clear \cite{compare}. Though discord is sometimes interpreted as a type of entanglement with nonclassical correlations, this interpretation is not accurate \cite{luo,ali}. In an open system, we also witnessed such a region where the entanglement is larger than quantum discord. For maximally entangled Bell state or in general any pure state, quantum mutual information is evenly distributed among classical and quantum correlations \cite{ali,luo}. We found that for the case of a system with two interacting fermions, the quantum discord of a steady state is always larger than the classical correlation, which is not true in general \cite{371}. In the high temperature regime, the quantum discord and classical correlation coincide and exponentially decrease as temperature increases.

In this paper, we study a fermionic system coupled with two separate bosonic or fermionic reservoirs. We analyze the influence of nonequilibriumness to quantum correlations and separate the correlation generation due to \textit{essential nonequilibriumness} from that arises from the \textit{averaged effect} of the two baths. We show that the distilled nonequilibrium effect only \textit{enhances} the quantum correlations, while averaged effect can have varying influences on correlations. We notice that in the large tunneling rate regime with fermionic environments, quantum correlations do not necessarily vanishes as we enlarge the chemical potential of one of the reservoirs. Some threshold value emerges above which the asymptotic entanglement remains finite, otherwise zero. In some parameter regime, quantum entanglement \textit{resurrects} after previous dying-out with the increase of chemical potential bias. 

The paper is organized as follows. In section II, we introduce the model and the quantum master equation. In section III, we briefly introduce different quantum correlations that are relevant and the necessary analytical calculations. In section IV, we analyze the quantum correlations in both bosonic and fermionic reservoirs when the system is in equilibrium with the environment. The main point of the paper is made in section V and VI, where we discuss the results for the \textit{nonequilibrium effects on quantum correlations} and other nonequilibrium phenomena. We separate the equilibrium and nonequilibrium components of correlation generation and remark on the applicability of Lindblad form in understanding nonequilibrium physics and the different behaviors of correlations at various tunneling rates. In the last section, we conclude our results.

\section{Model}

\begin{figure}[ht]
\centering
\includegraphics[width=0.44\textwidth]{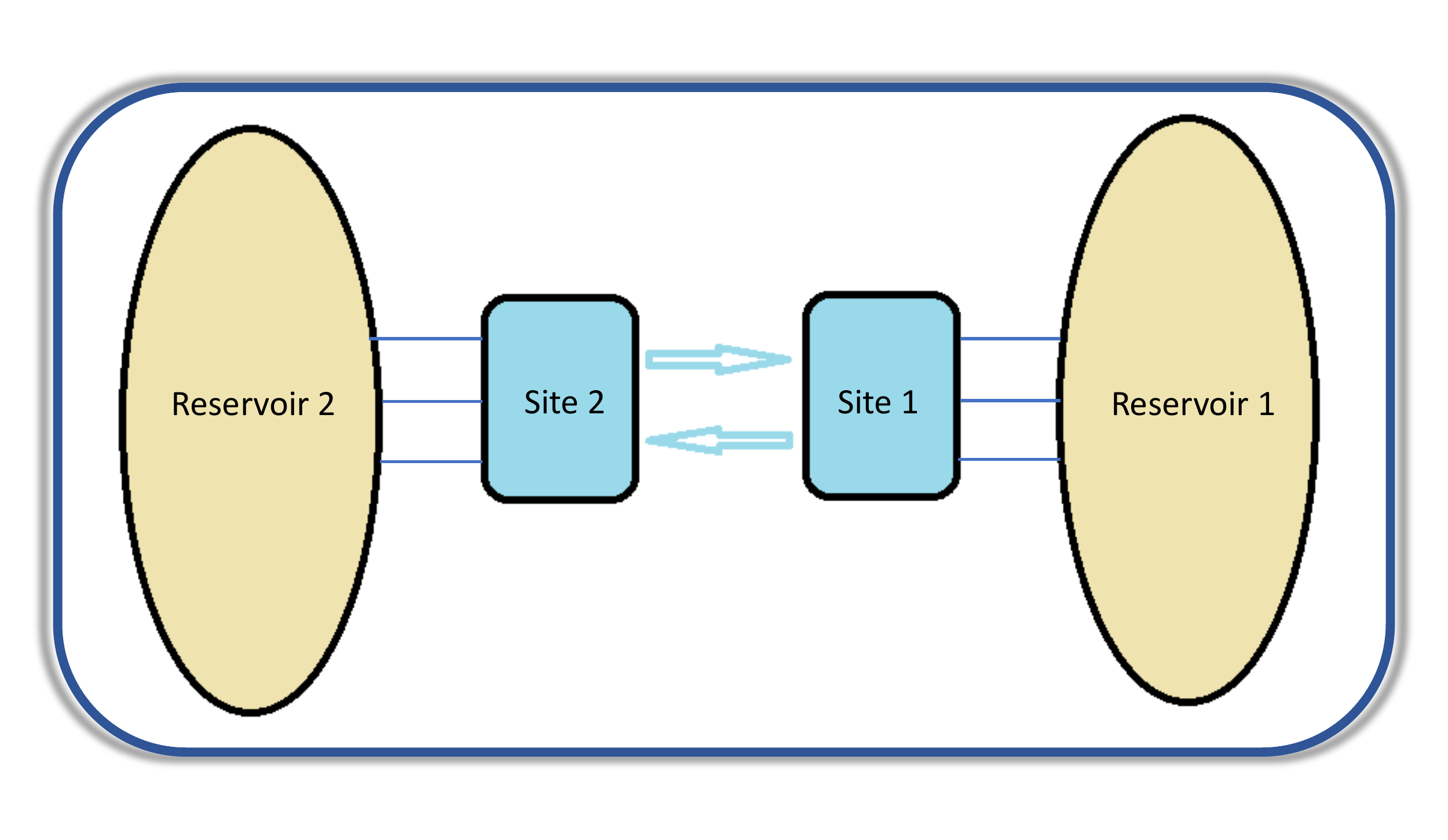}
\caption{Schematic illustration of the model under consideration. The two fermion sites are either occupied or empty with nonzero tunneling rate between them. Each site is in contact with its own reservoir.}
\label{figmain}\end{figure}

We consider a simplest model with two sites. Each site can either adopt a fermon or be empty. The fermion can tunnel through between the two sites with a finite tunneling rate. Each site is immersed in its own reservoir which can be either bosonic with zero chemical potential or fermionic. The diagrammatic illustration of our model is shown in Fig.~\ref{figmain}. We take the Hamiltonian of the following form:
\beq \bsplit
&\calh_S=\omega_1 \eta_1^\dagger \eta_1+\omega_2 \eta_2^\dagger \eta_2+\Delta(\eta_1^\dagger \eta_2+ \eta_2^\dagger \eta_1) \\
&\calh_R=\sum_{k,p}\hbar \omega_k \ (a_{kp}^\dagger a_{kp})+\sum_{q,s}\hbar \omega_q \ (b_{qs}^\dagger b_{qs})
 \end{split}\eeq
where $\calh_S$ and $\calh_R$ represent the hamiltonian of the system and the hamiltonian for the reservoirs, respectively. $\Delta$ is the interaction strength between the two sites. $\eta_{1,2}^\dg$ and $\eta_{1,2}$ are creation and annihilation operators on the site 1(2) following the standard fermionic statistics
\beq \begin{split}& \{\eta_a,\eta_b^\dg\}=\delta_{ab}, \\  &\{\eta_a,\eta_b\}= \{\eta_a^\dg,\eta_b^\dg\}=0. \end{split} \label{fermi}\eeq

The interaction Hamiltonian between the system and the reservoirs is denoted as
\beq \mathcal{H}_{int}=\sum_{k,p}\lambda_k \ (\eta_1^\dagger a_{kp}+ \eta_1 a^\dagger_{kp})+\sum_{q,s} \lambda_q \ (\eta_2^\dagger b_{qs}+ \eta_2 b^\dagger_{qs}) \eeq
with $\lambda$ being the interaction strength between the system and the reservoir and $a_{kp}^{\dg} (b_{kp}^\dg)$ the creation operator for a particle of momentum $k$, polarization $p$ from the reservoir. For bosonic reservoirs, the calculation serves as a toy model for energy and heat transport \cite{transport1} since it is not possible from the first principle. For fermionic reservoirs, this model can be used to describe charge transport \cite{transport2}.
We can diagonalize the Hamiltonian with the following transformation, \beq \vec{\zeta}=\begin{pmatrix} \cos(\theta/2) & \sin(\theta/2) \\ -\sin(\theta/2) & \cos(\theta/2) \end{pmatrix} \vec{\eta} \label{newH}\eeq
where cos$\theta=\dfrac{\omega_2-\omega_1}{\sqrt{(\omega_1-\omega_2)^2+4 \Delta^2}}$. After the diagonalization, \beq \calh_S=\omega'_1 \zeta_1^\dagger \zeta_1+\omega'_2 \zeta_2^\dagger \zeta_2 \eeq with $\omega'_{1,2}=\frac{1}{2} (\omega_1+\omega_2 \pm \sqrt{(\omega_1-\omega_2)^2+4 \Delta^2})$ and $\zeta_{1,2}$. $\zeta_{1,2}^\dg$ satisfy the same anti-commutation relation as \ref{fermi}. After this Bogoliubov transformation, this model effectively describes two uncoupled fermion sites with the energies to generate one of the fermions $\omega_1^\prime$ and another fermion $\omega_2^\prime$, respectively. Bogoliubov transformation does not change the commutation relation. In the new basis, $\zeta_a$ and $\zeta^\dagger_b$ still have the same commutation relation as the $\eta_a$ and $\eta_b^\dagger$. The interaction Hamiltonian can be written as
\beq \calh_{int}=\sum_{k,p} \lambda_k((s\zeta_1^\dg+c \zeta_2^\dg)a_{kp}+(c\zeta_1^\dg-s\zeta_2^\dg)b_{kp})+h.c., \eeq
where $c= \cos(\theta/2)$ and $s= \sin(\theta/2)$.

\subsection{Master equation}
With the Born approximation, the density matrix $\rho_{SR}$ for the whole system can be factorized into the following form when the interaction between the systems and environment are assumed to be weak, i.e. $\tilde \rho_{SR}(t)\approx \tilde \rho_S(t) \otimes \tilde \rho_R$,
where $\tilde \rho_S(t)$ is the density matrix for the isolated system and $\tilde \rho_R$ for the isolated reservoirs. We ignore the variance in the reservoirs, i.e. $\til \rho_R(t) \approx \rho_R(0)=\rho_R$. When the Markovian approximation is assumed, the quantum master equation (QME) can be written as follows \cite{breuer}
\beq \dfrac{d \tilde \rho_S(t)}{dt} =-\frac{1}{\hbar^2}\mathrm{Tr}_R \int_0^t ds \left[\tilde H_{int}(t),[\tilde H_{int}(s),\til \rho_S(t)\otimes \rho_R(0)]\right] \eeq
where $\til{H}_{int}\equiv e^{iH_0t/\hbar}H_{int}e^{-iH_0t/\hbar}$ is the interaction Hamiltonian in the interaction picture, $H_{int}$ is the interaction Hamiltonian in the Schr\"odinger picture and $H_0=H_S+H_R$. 

The environment in our consideration is macroscopic and has a much shorter relaxation time compared to the system's characteristic time scale. Thus we ignore the back reaction from the reservoir to our system. We trace out the environmental contributions to the full density matrix to obtain the reduced density matrix of our system. For an ideal bosonic or a fermionic environment, the following results hold \beq \begin{split} &Tr_R (\rho_R \ a_k^\dg  a_s)=\delta_{ks} n_k \\ & Tr_R (\rho_R \ a_k a_s^\dg)=\delta_{ks}(1\pm n_k)  \\ &Tr_R (\rho_R \ a_k a_s)=Tr_R (\rho_R \ a_k^\dg a_s^\dg)=0,\end{split} \eeq where the plus sign is for bosonic bath with occupation number $n_k=\dfrac{1}{e^{\hbar \beta \omega}-1}$, and the minus sign is for fermioic bath with $n_k=\dfrac{1}{e^{\hbar \beta(\omega_k-\mu)}+1}$.

One of the most typical ways of dealing with quantum open system is to apply Lindblad operator and solve the corresponding master equation. However, the master equations in the Lindblad form ignore certain population and coherence couplings and assume extra time scale hierarchy that is not widely applicable. The coherence and population couplings are important for the system in the nonequilibrium environments \cite{prl18}. In our study, we will apply Markovian approximation without secular approximation. This gives arise to the Bloch-Redfield equation. The Redfield equation may have the problem of positivity in principle \cite{positivity1}, but it does not appear within any reasonable range of parameters in our calculation. In our study, we restrict to the parameters within the reasonable range where the density matrix is positive definite. The result of the quantum master equation is displayed as follows,
\beq \dot \rho_S(t)=i[\rho_S,H_S]-D_0[\rho]-D_s[\rho], \label{mst} \eeq
where
\beq D_0[\rho]=\sum_{i=1}^2N_i[\rho], \ \ D_s[\rho]=\sum_{i=1}^2 S_i[\rho]. \eeq
The dissipator $D_0$ describes the particle exchanges with the reservoirs, and $D_s$ gives the coherence between energy levels of the system which is absent in the Lindblad formalism. For bosonic reservoirs, $D_0$ and $D_s$ are defined as follows,
\begin{widetext}
\begin{align} N_1[\rho]&=\Gamma_1 \cdot s^2 \ [(1\pm n_1^{T_1})(\zeta_1^\dagger \zeta_1 \tilde \rho + \rhot \zeta_1^\dagger \zeta_1-2 \zeta_1 \rhot \zeta_1^\dagger) +n_1^{T_1} (\zeta_1 \zeta_1^\dagger \rhot+\rhot \zeta_1 \zeta_1^\dagger-2\zeta_1^\dagger \rhot \zeta_1)] \nonumber\\
&+\Gamma_2 \cdot c^2 \ [(1\pm n_2^{T_1})(\zeta_2^\dagger \zeta_2 \tilde \rho + \rhot \zeta_2^\dagger \zeta_2-2 \zeta_2 \rhot \zeta_2^\dagger) +n_2^{T_1} (\zeta_2 \zeta_2^\dagger \rhot+\rhot \zeta_2 \zeta_2^\dagger-2\zeta_2^\dagger \rhot \zeta_2)],\end{align}
and
\begin{align}
S_1[\rho]&=\Gamma_1\cdot s\cdot c \ [(1 \pm n_1^{T_1})(\zeta_2^\dagger \zeta_1 \rhot+ \rhot \zeta_1^\dagger \zeta_2  - \zeta_2 \rhot \zeta_1^\dagger- \zeta_1 \rhot \zeta_2^\dagger) + n_1^{T_1} (\zeta_2 \zeta_1^\dagger \rhot+\rhot \zeta_1 \zeta_2^\dagger- \zeta_2^\dagger \rhot \zeta_1 - \zeta_1^\dagger \rhot \zeta_2 )] \nonumber\\
& + \Gamma_2\cdot s\cdot c \ [(1 \pm n_2^{T_1})(\zeta_1^\dagger \zeta_2 \rhot + \rhot \zeta_2^\dagger \zeta_1  - \zeta_1 \rhot \zeta_2^\dagger - \zeta_2 \rhot \zeta_1^\dagger) +n_2^{T_1} (\zeta_1 \zeta_2^\dagger \rhot +\rhot \zeta_2 \zeta_1^\dagger - \zeta_1^\dagger \rhot \zeta_2 - \zeta_2^\dagger \rhot \zeta_1 )].
\end{align} \end{widetext}
where $\omega_a'$ is the energy eigenvalue of the system and $s$ and $c$ are the short from for sin$(\theta/2)$ and cos$(\theta/2)$ defined in \ref{newH}. Plus signs are for bosonic reservoirs and minus signs for fermionic reservoirs. Due to the rapid oscillation of field modes,  we apply the Weisskopf-Wigner approximation, expand the time integral to infinity and replace the summation in the interaction Hamiltonian by integration. Then, the decay rates are defined as
\beq \Gamma_i \equiv \frac{2V}{(2\pi)^3}\pi\int d^3\vec{k}\ \lambda^2_{\vec{k}}\ \delta(\omega'_i-\omega_k) \eeq
The number density is defined as follows. For the bosonic bath, $n_k^{T_i}=\dfrac{1}{e^{\hbar \beta_i\omega'_k}-1}$, and for fermionic bath $n_k^{T_i}=\dfrac{1}{e^{\beta_i(\omega'_k-\mu_i)}+1}$. The $N_2[\rho]$ and $S_2[\rho]$ differ from $N_1[\rho]$ and $S_1[\rho]$ by replacing the $T_1$ in the above expressions with $T_2$, the $c$ with $-s$ and $s$ with $c$. 

The solution of the master equation has two uncoupled parts, $\rho_{11},\ \rho_{22},\ \rho_{33},\ \rho_{44},\ \rho_{23},\ \rho_{32}$ and the rest. The off-diagonal components, except $\rho_{23}$ and $\rho_{32}$, are uncoupled with the population components, and thus they vanish in the steady state. Therefore, we only consider the "X" form of the density matrix
\bea  \rho=\begin{pmatrix}\rho_{11} & 0 & 0 & 0 \\ 0& \rho_{22} & \rho_{23} & 0 \\ 0&\rho_{32}& \rho_{33}&0\\0&0&0&\rho_{44} \end{pmatrix}. \eea
The solution of the master equation is in the energy representation. We need to perform a unitary transformation to transform the density matrix to the local site basis to calculate the correlations between two sites. The explicit form is given in the Appendix \ref{elements}.

\subsection{Eigenstates and connection with the spin operator}
\subsubsection{Relation to spin $\frac{1}{2}$}
Spin-$\frac{1}{2}$ system and fermionic system can be mapped into one another through Jordan-Wigner transformation. For a fermionic system, i.e. two holes which admit either zero or one fermion, the original Hamiltonian can be presented in the language of spin system. We take the local basis $ |0\rangle\otimes |0\rangle,\  |0\rangle\otimes |1\rangle,\  |1\rangle\otimes |0 \rangle, \ |1\rangle\otimes |1\rangle $. Correspondingly, the creation and the annihilation operators take the form
\beq \begin{split} \eta_1^\dagger &= \sigma^+ \otimes \mathbbm{1}, \hspace{0.6cm} \eta_2^{\dagger} =\sigma_z \otimes \sigma^+, \\  \eta_1 &= \sigma^- \otimes \mathbbm{1}, \hspace{0.6cm} \eta_2 =\sigma_z \otimes \sigma^-, \end{split} \label{2.10} \eeq
where  $\sigma^+$ and $\sigma^-$ are the raising and lowering operators for the spin system, $\sigma_z$ is the Pauli spin matrix. The creation and annihilation operators follow the fermionic anticommutation relation,
\beq \{\eta_a,\  \eta^\dagger_b\}= \delta_{ab}, \hspace{0.6cm} \{\eta_a,\  \eta_b\}= \{\eta_a^\dagger,\  \eta^\dagger_b\}=0.\eeq

The system Hamiltonian thus can be transformed into the following form
 \begin{multline} \calh_S=\omega_1 \ \sigma^+ \sigma^- \otimes \mathbbm{1}  +\omega_2 \ \mathbbm{1} \otimes \sigma^+\sigma^- \\ +\Delta(\sigma^+ \sigma_z \otimes \sigma^- + \sigma_z \sigma^- \otimes \sigma^+), \end{multline}
and the interaction Hamiltonian can be written out accordingly,
\beq \calh_{int}=\sum_{k,p} \lambda_k(\sig^+\otimes \mathbbm{1} \ \otimes a_{k,p}+\sig_z \otimes \sig^+\otimes \ b_{k,p})+h.c. \eeq

\subsubsection{Eigenbasis}
We set our notation for basis in the following way. The eigenstate corresponding to zero energy, denoted by $|00\rangle$, is the one which can be annihilated by both $\zeta_1$ and $\zeta_2$.  The highest energy eigenstate corresponding to $w'_1+w'_2$, denoted by $|11\rangle$, is the one that can be annihilated by both $\zeta_1^\dagger$ and $\zeta_2^\dagger$.  In the local basis, they are identified under the above requirements to the states as $|0\rangle\otimes |0\rangle$ and $|1\rangle\otimes |1\rangle$, respectively. Applying \eqref{2.10}, we now have our four eigenstates defined as follows
\beq \begin{split}&|1\rg=\zeta_1|10\rg=\zeta_2|01\rg=|0\rg \otimes|0\rg \\ & |2\rangle=\zeta_1^\dagger |00\rg =\zeta_2 |11\rg =c \ |1\rangle\otimes|0\rangle -s \ |0\rangle\otimes|1\rangle  \\ & |3\rg= \zeta_2^\dg |00 \rg=-\zeta_1|11\rg = -s\ |0\rg\otimes |1\rg -c\ |1\rg\otimes |0\rg\\ & |4\rg=-\zeta_1^\dg |01\rg=\zeta_2^\dg |10\rg=|1\rg\otimes |1\rg, \end{split} \label{basis} \eeq
and with eigenvalues $0, \omega_1',\ \omega_2',\ \omega_1+\omega_2$. Notice that the ground state and the most excited eigenstate in our definition are localized, while the other two are nonlocal.

For simplicity, we consider the symmetric case $\omega_1=\omega_2=\omega$, and with the following transformation
\beq \vec{\zeta}=\begin{pmatrix} \dfrac{1}{\sqrt{2}} & \dfrac{1}{\sqrt{2}} \\ -\dfrac{1}{\sqrt{2}} & \dfrac{1}{\sqrt{2}} \end{pmatrix} \vec{\eta}, \eeq
the Hamiltonian can be simplified to
\beq \calh_S=(\omega + \Delta)\zeta_1^\dg \zeta_1 +(\omega-\Delta) \zeta_2^\dg \zeta_2  \label{ham}. \eeq
Without consideration of the system-reservoir interaction, the eigenstate annihilated by $\zeta_2$ has energy $\omega+\Delta$ and the state annihilated by $\zeta_1$ has energy $\omega-\Delta$.  Since the four local states $ |0\rangle\otimes |0\rangle,\  |0\rangle\otimes |1\rangle,\  |1\rangle\otimes |0 \rangle, \ |1\rangle\otimes |1\rangle$ form a complete basis set, the energy eigenbasis can be expressed in terms of the four state vectors. The energy eigenstate can be written as a linear combination of the four states and solved by requiring that $\calh_S\ |n'\rg =\omega_n'\ |n' \rg$. The eigenbasis and the corresponding energies are listed as follows, 
\begin{align}
&|0\rangle\otimes |0\rangle   &&E_1=0\\
&\frac{1}{\sqrt 2}(|0\rg\otimes |1\rg+|1\rg\otimes|0\rg) \quad  &&E_2=\omega'_1=\omega-\Delta \\
&\frac{1}{\sqrt 2}(|0\rg\otimes |1\rg-|1\rg\otimes|0\rg) \quad &&E_3=\omega'_2=\omega+\Delta\\
&|1\rg\otimes|1\rg   &&E_4=2\omega.
\end{align}

\numberwithin{equation}{section}
\section{Correlation measures of two fermions}
The feature of correlation of a bipartite system can be quantified by many physical quantities such as quantum mutual information, quantum discord, entanglement, as well as classical correlation. In this paper, we study the system correlation by these measures. We work in local basis and the subscript "local" will be neglected in the discussion of correlations in this chapter.

\subsection{Concurrence}

Concurrence was derived from the entanglement formation and is an entanglement measure that was introduced for describing two-qubit systems by Wootters \cite{wootters}. We will borrow this concept for our study since the density matrix has the same dimension. There are several related studies on concurrence for e.g. two qubit system, three-level system \cite{wuwei1, cui, zd}.  The concurrence for a four dimensional density matrix (for example, two-qubit systems) is given by
\beq \mathcal{E(\rho)}=\text{Max}(0,\sqrt{\lambda_1}-\sqrt{\lambda_2}-\sqrt{\lambda_3}-\sqrt{\lambda_4)}\eeq
where $\lambda_i$ are the eigenvalues of the Hermitian matrix R in the descending order where
\beq R=\rho(\sigma_y\otimes\sigma_y)\rho^*(\sigma_y\otimes\sigma_y).\eeq
For the "X" type density matrix that is related to our study, it can be easily calculated that the concurrence reduces to the following simple expression,
\beq \mathcal{E(\rho)}=2\ \text{Max}(0,|\rho_{23}|-\sqrt{\rho_{11}\rho_{44}},|\rho_{14}|-\sqrt{\rho_{22}\rho_{33}}). \eeq

\subsection{Quantum mutual information and classical correlation}
Quantum mutual information (QMI) is a direct generalization of classical mutual information and is defined as follows. Let $\rho^{AB}$ denotes the density operator of a bipartite system $AB$, and $\rho^{A(B)}=Tr_{B(A)}(\rho^{AB})$ is the reduced density operator of the subsystem $A$($B$), respectively. Then the QMI can be expressed as
\begin{eqnarray}
\mathcal{I} (\rho^{AB}) = S (\rho^A) + S (\rho^B) - S(\rho^{AB})\ , \label{QMI}
\end{eqnarray}
where $S(\rho) = - \mathrm{tr} \, ( \rho \, \log_2 \rho )$ is the von Neumann entropy. It is shown that quantum mutual information is the maximum amount of information that A can send securely to B if A and B share a correlated composite quantum system and AB is used as the key for a one-time pad cryptography system \cite{qmi1}.

Another generalization of classical mutual information is classical correlation (CC) and is defined as below. Let ${B_k}$ be a set of one-dimensional projection measurement performed on subsystem $B$, the conditional density operator $\rho_k$ associated with the measurement result $k$ is
 \begin{eqnarray}
\rho_k=\frac{1}{p_k}(I\otimes B_k)\rho (I\otimes B_k),
\end{eqnarray}
where $p_k=tr(I\otimes B_k)\rho (I\otimes B_k)$, $I$ is the identity operator on the subsystem $A$.

The von Neumann measurement for subsystem $B$ can be written as \cite{luo}
\begin{eqnarray}
B_i = V \, \Pi_i \, V^\dagger : \quad i = 0,1
\end{eqnarray}
where $\prod_i=|i\rangle\langle i|$ is the projector associated with the subsystem $B$ along the computational basis $|i\rangle$, and $V \in SU(2)$ is a unitary operator.

With this conditional density operator, the quantum conditional entropy with respect to this measurement is
defined by
\begin{eqnarray}
S(\rho|\{B_k\}):=\sum_{k}p_kS(\rho_k), \label{cond}
\end{eqnarray}
and the associated quantum mutual information is given by
\begin{eqnarray}
\mathcal{I}(\rho|\{B_k \}):=S(\rho^A)-S(\rho|\{B_k\}).
\end{eqnarray}
Classical correlation is defined as the superior of $\mathcal{I}(\rho|\{B_k\})$ over
all possible von Neumann measurement ${B_k}$ \cite{cc},
\begin{eqnarray}
\mathcal{C}(\rho):=\sup_{\{B_k\}}I(\rho|\{B_k\}).
\end{eqnarray}

\subsection{Quantum discord}
Quantum discord $\mathcal{Q}$, which reflects the quantum correlations between the two subsystems, was introduced by Ollivier and Zurek in 2001 \cite{ollivier}. They found that even for a separable state, a measurement on the subsystem can still disturb the whole system unless $\mathcal{Q}$=0. 

The sum of quantum discord $\mathcal{Q}(\rho)$ and classical correlation $\mathcal{C}(\rho)$ is the quantum mutual information  $\mathcal{I}(\rho)$, i.e.
\begin{eqnarray}
\mathcal{Q}(\rho):=\mathcal{I}(\rho)-\mathcal{C}(\rho). \label{disc}
\end{eqnarray}
The quantum discord in general is not symmetric regarding to which system is the operator performed upon, i.e. $\mathcal{I}(\rho|\{B_k \}) \ne \mathcal{I}(\rho|\{A_k \})$ even in their limits. In this article we will stick with the above definition and the conditional entropy is defined with respect to the measurement on the subsystem Site 2 as is shown in Fig \ref{figmain}.

The quantum discord of Bell-diagonal states is well-known \cite{horodecki2}. The analytical expressions for classical correlation and quantum discord are available for two-qubit Bell diagonal state and a seven-parameter family of two-qubit ``X" states \cite{luo,ali}. Though the analytical results for the seven-parameter family were pointed out not always exact, they are very good approximations in most cases \cite{errorqd,error1,error2}. For a general Hermitian operator acting on a $C^2 \otimes C^2$, the density matrix can be decomposed using the tensor products of $su(2)$ generators, i.e.
\beq
\rho=\frac{1}{4}[I\otimes I+\pmb{r \cdot \sigma} \otimes I+ I\otimes \pmb{s\cdot \sigma}+\sum_{i,j=1}^3c_{ij}\sigma_i\otimes\sigma_j]
\eeq
where coefficient $c_{ij} \in \mathcal{R}^3$.  For the class of ``X" state, the Bloch vector is along the z-axis, and the above expression can be simplified to
\beq
\rho=\frac{1}{4}[I\otimes I+r \cdot \sigma_z \otimes I+ I\otimes s \cdot \sigma_z+\sum_{i=1}^3c_{ij}\sigma_i\otimes\sigma_j] \label{tensor}
\eeq
with $c_{13}=c_{23}=c_{31}=c_{32}=0$. Quantum discord is invariant under the local unitary transformations. It can be easily shown that every general 2$\times$2 state of the form \ref{tensor} can be reduced to a form of which the coefficients $c_{ij} \in \mathcal{R}^3$are diagonal by a local unitary transformation, i.e.,
\begin{theorem}
$\forall$ $c_{ij} \in \mathbf{R}$, there exists unitary matrices $U$ and $V$ such that $U\otimes V(\sum_{i,j=1}^3c_{ij}\sigma_i\otimes\sigma_j) U^\dg \otimes V^\dg=\sum_{m=1}^3 c_m\sig_m\otimes \sig_m$ for some $c_m$'s.\end{theorem}
The proof of the above statement can be found in many literature (e.g. \cite{luo}) and we will not elaborate here.

\subsection{Quantum correlations in our case of study}

In the case of our study, the explicit calculation for quantum mutual information, quantum discord and classical correlation is given below. In order to diagonalize the coefficients $c_{ij}$, we define the following unitary transformations,
\begin{eqnarray}
&\mathcal{L}_\theta (\sigma_x)& \equiv e^{i\theta\sigma_z} \sigma_x  e^{-i\theta\sigma_z}=\cos{2\theta}\ \sigma_x-\sin{2\theta}\  \sigma_y \nonumber \\
&\mathcal{L}_\theta (\sigma_y)& \equiv e^{i\theta\sigma_z} \sigma_y e^{-i\theta\sigma_z} =\sin{2\theta}\ \sigma_x + \cos{2\theta}\ \sigma_y \nonumber \\
&\mathcal{L}_\theta (\sigma_y)& \equiv e^{i\theta\sigma_z} \sigma_z e^{-i\theta\sigma_z} = \sigma_z
\end{eqnarray}
where the parameters $\theta$ and $\psi$ solve the following equation
\bea \mathcal{L}_\theta \otimes \mathcal{L}_\psi (\sum_{i=1}^3c_{ij}\sigma_i\otimes\sigma_j) =(\sum_{i=1}^3 g_{i}\sigma_i\otimes\sigma_i).\eea
The solution is not unique and we pick the  simplest one for the calculation
\beq \cos(2\theta)=\dfrac{\Im(\rho_{23})}{|\rho_{23}|}, \quad \sin(2\theta)=\frac{\Re(\rho_{23})}{|\rho_{23}|}, \eeq
where $\Im(\rho_{ij})$ denotes the imaginary part of the matrix element $\rho_{ij}$ and $\Re(\rho_{ij})$ denotes the real part. After the transformation, the density matrix in the local basis representation takes the following form, i.e.
\bea
\rho=\begin{pmatrix} \rho_{11} & 0 & 0 & 0 \\ 0& \rho_{22} & |\rho_{23}| & 0 \\ 0&|\rho_{32}|& \rho_{33}&0\\0&0&0&\rho_{44}. \label{x} \end{pmatrix}
\eea
The density matrix belongs to real-valued four-parameter family, and it can be decomposed in the following form,
\begin{multline}
\rho=\frac{1}{4}[I\otimes I+r \cdot \sigma_z \otimes I+s\cdot I\otimes s\sigma_z+\sum_{i=1}^3c_{i}\sigma_i\otimes\sigma_i]. \label{tensr}
\end{multline}
where $r,\ s$ and $c_3$ can be solved as follows,
\beq \begin{split} &s=(\rho_{11}-\rho_{22})+(\rho_{33}-\rho_{44}) \\ &r=(\rho_{11}+\rho_{22})-(\rho_{33}+\rho_{44}) \\ &c_1=c_2=2|\rho_{23}| \\  &c_3= (\rho_{11}-\rho_{22})-(\rho_{33}-\rho_{44}). \end{split} \eeq
The eigenvalues of the density matrix $\rho$ are
$$
\begin{array}{l}
\lambda_{1,2}=\frac{1}{4}[1-c_3\pm\sqrt{(r-s)^2+(c_1+c_2)^2} ],\\[2.5mm]
\lambda_{3,4}=\frac{1}{4}[1+c_3\pm\sqrt{(r+s)^2+(c_1-c_2)^2} ]\,.
\end{array}
$$
and the quantum mutual information is given as follows,
\begin{eqnarray}
\mathcal{I}(\rho)&=&S(\rho^A)+S(\rho^B)+\sum_{i=1}^{4}\lambda_i\log_2 \lambda_i \label{mutualinformation}
\end{eqnarray}
where $S(\rho^A)$ and $S(\rho^B)$ are given by $S(\rho^A)=1+f(r)$, $S(\rho^B)=1+f(s)$ and $f(t)=-\frac{1-t}{2}\log_2(1-t)-\frac{1+t}{2}\log_2(1+t)$.

The quantum discord for the density matrix of the above form was first studied analytically \cite{luo,ali} and latter was explicitly exemplified \cite{Fei}. In most cases, it turns out that the possible maximal values in the optimization process of calculating classical correlation only take place in several special places with little error \cite{errorqd} and the result is as follows. For any state $\rho$ of the form \ref{x}, the classical correlation of $\rho$ is given by
\begin{eqnarray}
\label{proposition1}
\mathcal{C}(\rho)= S(\rho^A)- \min\{S_1, S_2\},
\end{eqnarray}where
\begin{eqnarray}
S_1&=& S(\rho|\{B_i\})=p_0S(\rho_0)+p_1S(\rho_1) \nonumber\\
   &=& -\frac{1+r+s+c_3}{4}\log_2\frac{1+r+s+c_3}{2(1+s)}\nonumber\\
    &-&\frac{1-r+s-c_3}{4}\log_2\frac{1-r+s-c_3}{2(1+s)}\nonumber\\
     &-&\frac{1+r-s-c_3}{4}\log_2\frac{1+r-s-c_3}{2(1-s)}\nonumber\\
      &-&\frac{1-r-s+c_3}{4}\log_2\frac{1-r-s+c_3}{2(1-s)},
\end{eqnarray}and
\begin{eqnarray}
S_2=1+f(\sqrt{r^2+c_1^2}),
\end{eqnarray}
where $f(t)=-\frac{1-t}{2}\log_2(1-t)-\frac{1+t}{2}\log_2(1+t)$. 
And the quantum discord is given by
\begin{eqnarray}
\mathcal{Q}(\rho)=\mathcal{I}(\rho)-\mathcal{C}(\rho).
\end{eqnarray}


\section{Quantum correlations for fermionic system in equilibrium environments}
The quantum correlations under equilibrium condition is important since the mechanism of correlation generation in the latter case can be the same as or different from that in the equilibrium scenario. This is what we want to differentiate. Under the equilibrium condition  the two reservoirs are set to the same temperature and the same chemical potential. At the long time limit, the system will relax to the same equilibrium regardless of the initial conditions. What we find is that all the four quantum correlations we study (QMI, CC, QD, concurrence) show the similar trends with respect to the temperature. As the temperature increases from zero, quantum correlations increase. As the temperature increases further, they decay asymptotically to zero, e.g. Fig. \ref{fig1}(a,b). While QMI, CC and QD decay asymptotically to zero, concurrence vanishes in an abrupt manner at T$=\Delta/\ln (1+\sqrt{2})$ regardless of the chemical potential (in fermionic reservoir case). In the fermionic reservoir case, quantum correlations show the same up-and-down trend with the increase of the temperature unless the chemical potential of the reservoirs $\mu$ is above $\mu^*=\omega-\Delta$ where they decay monotonically, see Fig. \ref{fig2}(b,c). The quantum correlations reach their maxima when chemical potential of the reservoirs and the eigenenergy of the system coincide.

Since our focus is on the nonequilibrium effects on the quantum correlations, we refer interested readers to the Appendix \ref{equilibrium} for details and full discussions on how equilibrium temperature and chemical potential influence the quantum correlations.

\section{Quantum correlations of fermionic system in nonequilibrium environments (Bosonic reservoirs)}
If we set the two reservoirs at different temperatures and chemical potentials, the system gradually evolves into the nonequilibrium steady state. The nonequilibrium steady state is featured by the constant flux of fermions between the two sites, and between the system and the reservoir. See Appendix \ref{current} for a more detailed description. For a microscopic system the relaxation to the steady state can happen extremely rapidly, and we will only focus on the steady state properties. Some of the previous studies have shown that non-equilbriumness can influence both the coherence and the entanglement of two qubits or a nanosystem in thermal or chemical reservoirs \cite{segal, wuwei1,zd, thermal}.

The section is arranged as follows. In subsection A, we briefly discuss the leading order solution of the Redfield equation and the inadequacy of the Lindbladian. In subsection B, we show that even in the nonequilibrium system, the effects on quantum correlations can still be divided into the equilibrium effect with the average of reservoirs, or "the averaged effect" and the nonequilibrium effect. We discuss the meaning of the averaged effect, how to separate this effect from the rest and we compare it with the "distilled" nonequilibrium effect.

\subsection{Leading order solution of Redfield equation}
The exact solution of the quantum master equation of the density matrix in bosonic baths is given in the Appendix \ref{sb}. Here, we present the solution up to the leading order in the ratio of system-reservoir interaction and tunneling rate. For the system with bosonic reservoirs,
\beq \begin{split} &\rho_{11}=\frac{(2 + n_{1p}) (2 + n_{2p})}{4 ((1 + n_{1p}) (1 + n_{2p})}+\mathcal{O}(g^2)\\
&\rho_{22}=\frac{n_{1p} (2 + n_{2p})}{4 (1 + n_{1p}) (1 + n_{2p})}+\mathcal{O}(g^2)\\
&\rho_{33}= \frac{(2 + n_{1p}) n_{2p}}{4 (1 + n_{1p}) (1 + n_{2p})}+\mathcal{O}(g^2)\\
&\rho_{44}=\frac{n_{1p} n_{2p}}{4 (1 + n_{1p}) (1 + n_{2p})}+\mathcal{O}(g^2)\\
&\rho_{23}=i\frac{n_{1m}(1+n_{2p}) + n_{2m}(1+n_{1p})}{2(1+n_{1p}) (1 + n_{2p})}g+\mathcal{O}(g^2),\end{split} \label{bs}\eeq
where $n_{ip}=n(\omega'_i, T_1) + n(\omega'_i, T_2)$, $n_{im}=n(\omega'_i, T_1) - n(\omega'_i, T_2)$ and $g=\frac{\Gamma}{\omega_1'-\omega_2'}=\frac{\Gamma}{2\Delta}$. It is valid the tunneling rate of the system $\Delta$ is much larger than the coupling between the system and the environment $\Gamma$.

For the nonequilibrium steady state, one of the most apparent changes is the appearance of the off-diagonal coherence terms in the density matrix of the system in the energy basis. This means that the steady state is not anymore "classical" in energy basis as is the case in equilibrium situation. This term was ignored by most previous studies but is important in studying nonequilibrium effect. 

One remark we make is that the coherence term $\rho_{23}$ is a distinctive feature of the \textit{Redfield} equation against \textit{Lindblad} equation. In the solution of the latter, the density matrix is a classical ensemble in the energy basis. If we want to study the real nonequilibrium effect, we have to abandon the rotating-wave approximation or the \textit{secular approximation}. The essential idea of secular approximation is that the system intrinsic time scale is much \textit{shorter} than the system's relaxation time, and thus we can ignore the unbalance of the system, in another words, ignore the nonequilibrium effect of the system. This point is elaborated in more detail in the fermionic reservoir section \ref{spcg}

\begin{figure}[ht]
\centering
\subfloat{{\includegraphics[width=0.33 \textwidth]{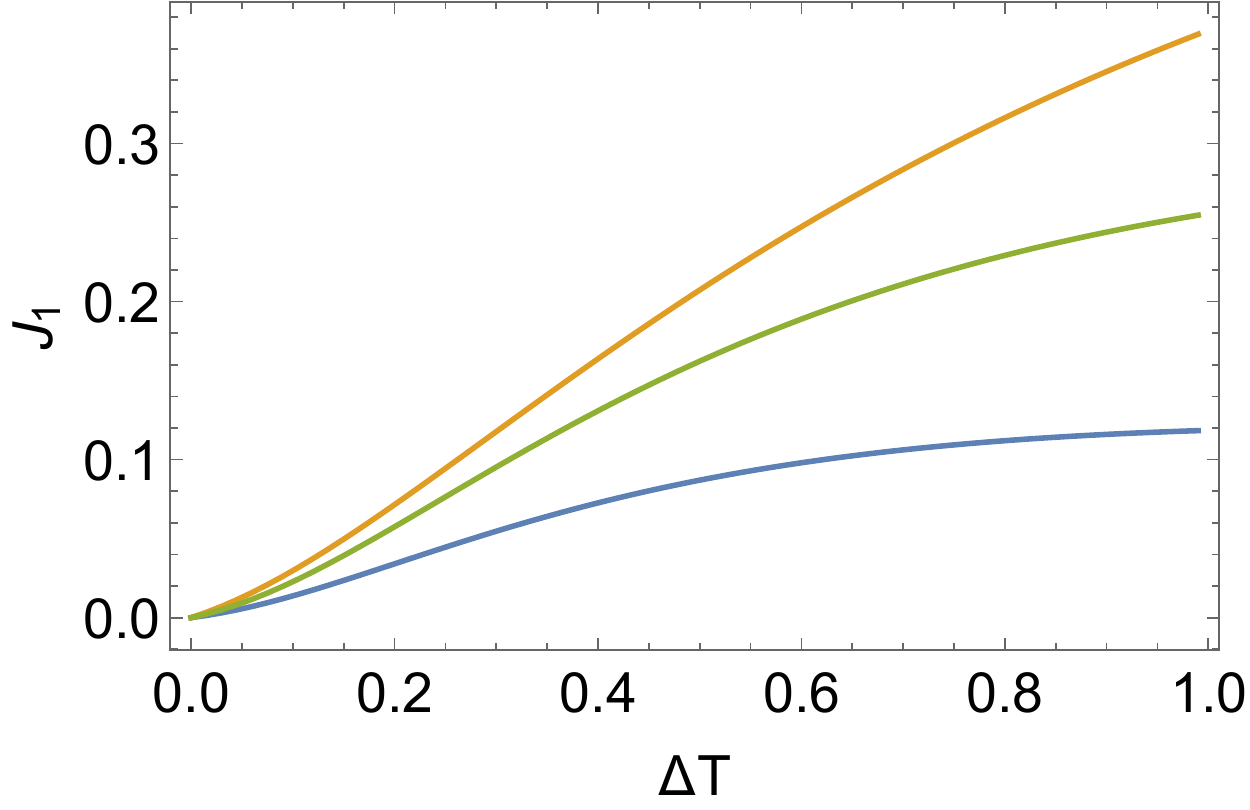}}}
\quad
\subfloat{{\includegraphics[width=0.34 \textwidth]{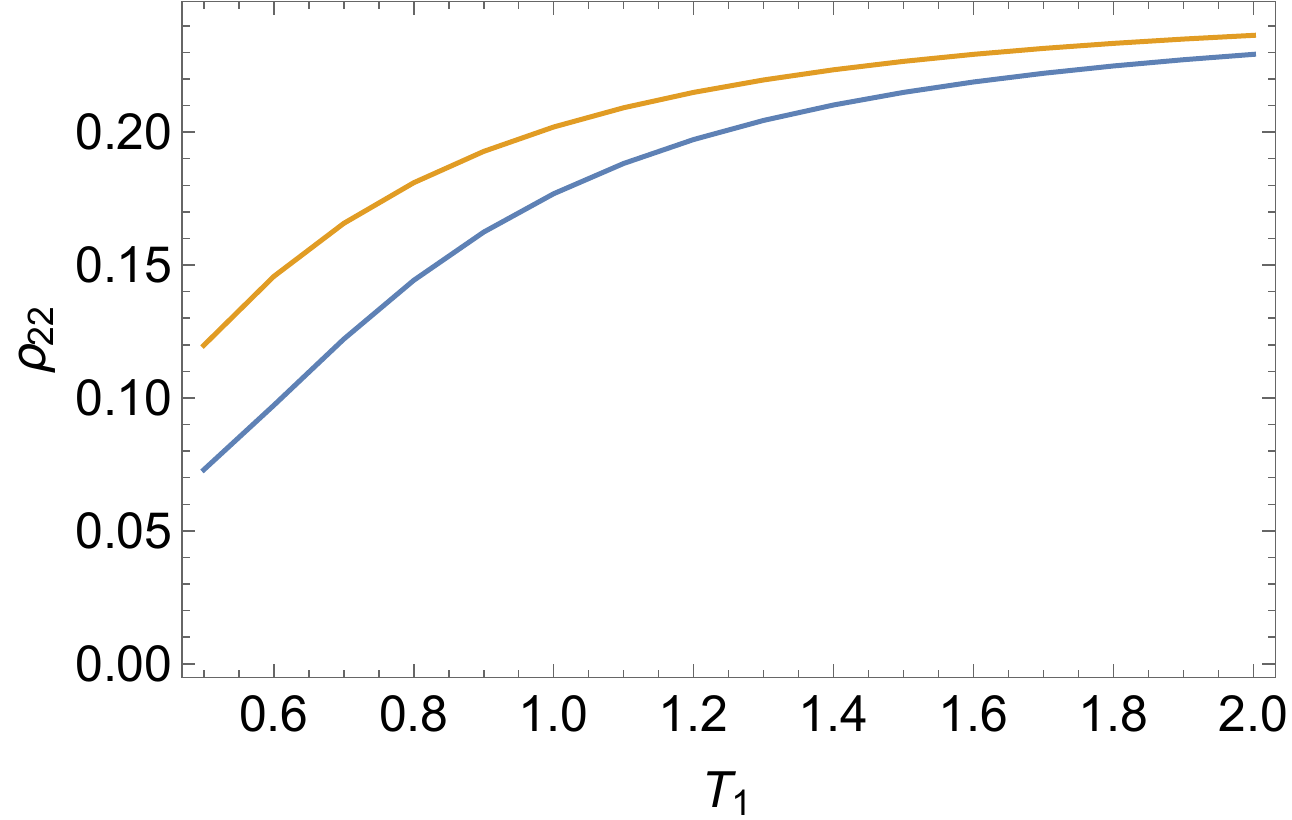}}}
\caption{(Color online)  (a) Energy current Vs $\Delta T$ at T$_1$=0.2. From top to bottom $\Delta=$0.3, 0.1, 0.05. (b) $\rho_{l,22}$  (lower, blue) and $\tilde \rho_{l,22}$  (upper, orange) Vs $T_1$ (the subscript "l" means local basis and the tilde indicates the equilibrium case) at $\Delta$T=0.4 and $\Delta=0.3$. For both, the parameters are set to $\Gamma_1=\Gamma_2=0.05$, $\omega_1=\omega_2=1$ unless otherwise specified.}
    \label{cb}
\end{figure}

\subsection{Nonequilibrium quantum correlations}
For both fermionic or bosonic environment, when setting the reservoirs (R1, R2) at the same chemical potential and different temperatures (say T$_1>$T$_2$), the expected particle number on site S1 is larger than S2. In terms of matrix elements, it means $\rho_{l,22}>\rho_{l,33}$ ($"l"$ represents "local basis"). However, comparing with equilibrium solution, the particle on site S1 is smaller than that in the equilibrium case with R1 at T$_1$, i.e. $\rho_{l,22}<\tilde{\rho}_{l,22}$, where $\tilde{\rho}_{l,22}$ is the local density matrix element in the equilibrium scenario at T$_1$, see Fig \ref{cb}(b). This means that comparing with the equilibrium case, the particles on site S1 has a constant net flow from S1 to S2 as is drawn in Fig \ref{cb}(a). Similarly, $\rho_{l,33}>\tilde{\rho}_{l,33}$, where $\tilde{\rho}_{l,33}$ is in the equilibrium case calculated at T$_2$.  S1 accepts fermions from the high temperature reservoir R1 and transports the fermions to S2 and then to R2. In steady state, the energy flowing into S1 has to be equal to that out from S2 to R2 in order to maintain the steady state. An introduction to energy current is given in Appendix \ref{current}. 

\begin{figure}[ht]
\centering
\subfloat[Large tunneling rate]{{\includegraphics[width=0.4 \textwidth]{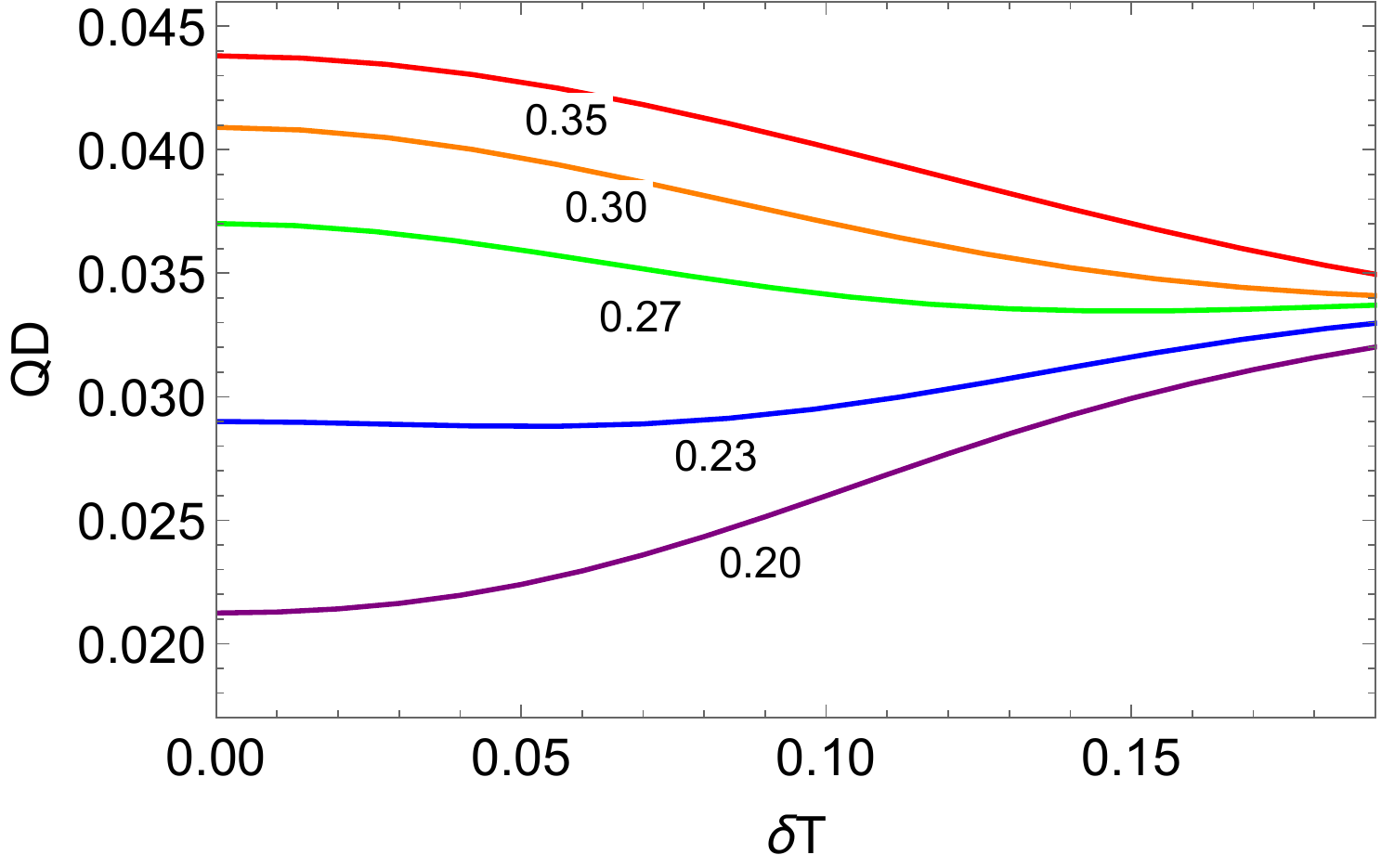}}}
\quad 
\subfloat[Small tunneling rate]{{\includegraphics[width=0.4 \textwidth]{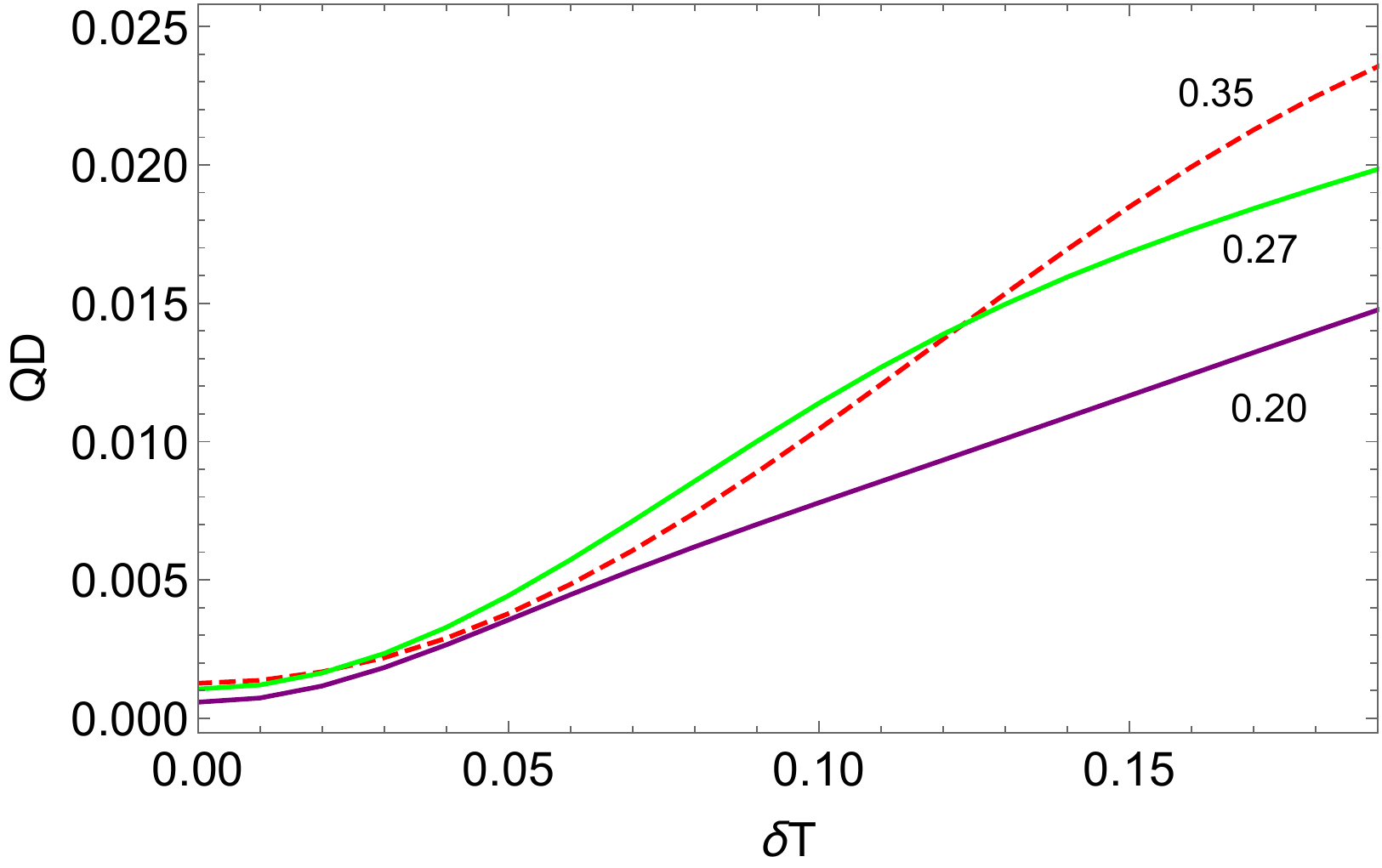}}}
\caption{(Color online) Nonequilibrium bosonic reservoirs. (a) The "rainbow" diagram. QD Vs $\delta$T with the average temperature T$_{avg}$ fixed at 0.2 (Purple), 0.23 (Blue), 0.27 (Green), 0.3 (Orange) and 0.35 (Red) from bottom to top. Here  $T_1=T_{avg}-\delta T$ and $T_2=T_{avg}+\delta T$. (b) QD Vs $\delta$T at $\Delta$=0.05 with different T$_{avg}$. The parameters are set to $\Delta=0.3$, $\Gamma_1=\Gamma_2=0.05$, $\omega_1=\omega_2=1$ unless otherwise specified.}
    \label{fig4}
\end{figure}

\subsubsection{Correlations due to the averaged effect}
When the tunneling rate is much larger than the system-environment coupling, the two subsystems which are in contact with their respective reservoirs act as one system in equilibrium with the average of the two reservoirs. When setting the $\Delta=0.3$ and $\Gamma=0.05$, we can ignore altogether the coherence and the higher orders in $g$. The quantum correlations should resemble the equilibrium case with the identification \ref{tbeff}. 
In the Fig \ref{fig4}(a), the "rainbow" diagram, we keep the average temperature of the two reservoirs fixed and plot the quantum discord against the change of the temperature bias of the two reservoirs, at the temperatures $T_{\text{avg}}\pm \delta T$, respectively. We can define an effective temperature $T_{eff}$ such that  
\beq n_i(T_{eff}) =\frac{n_i(T_{\text{avg}}+\delta T)+n_i(T_{\text{avg}}-\delta T)}{2}. \eeq 
We can check that the effective temperature
\beq T_{\text{i,eff}}=\omega'_i/\log{(\dfrac{2}{n_i(T_{\text{avg}}+\delta T)+n_i(T_{\text{avg}}-\delta T)}+1)}\label{Teff} \eeq
increases monotonically with the temperature bias $\delta T$. We notice at zero temperature bias, the system already has a sizable amount of discord. The increase or decrease of discord in the plot is mainly due to the average thermal temperature change of the two reservoirs. Through analysis of Fig \ref{fig1}, we concluded that the quantum discord increases with equilibrium temperature roughly when T$<0.3$ and decreases when T$>0.3$. Since $T_{eff}$ is $\delta T$ monotone, we expect the similar change of monotonicity at roughly T$_{avg}=0.3$ with $\delta T$. This is shown in Fig \ref{fig4}(a), the monotonicity of the discord with $\delta T$ changes when the average temperature is close to 0.3.

\begin{figure}[ht]
\includegraphics[width=0.4 \textwidth]{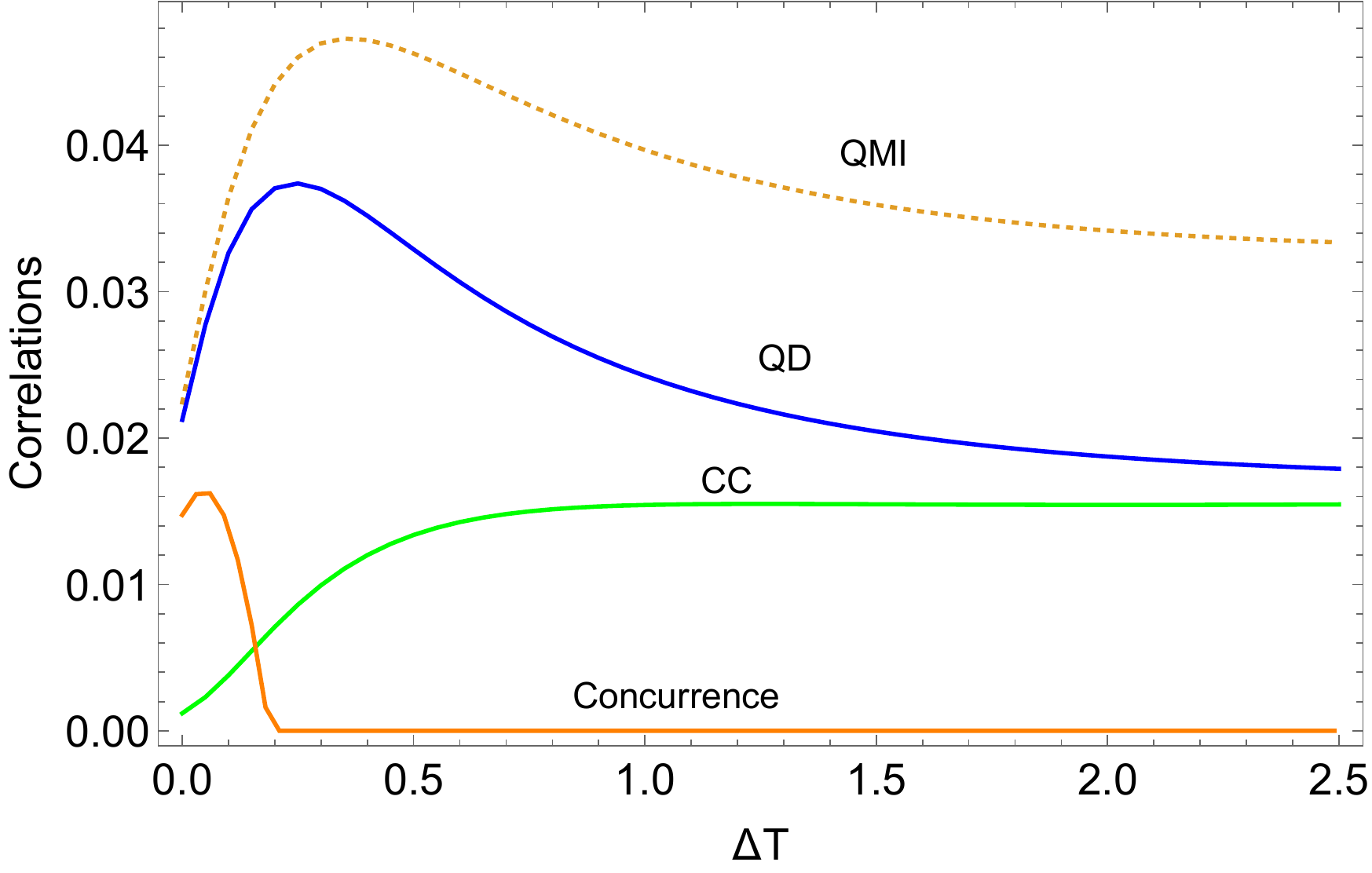}
\caption{(Color online) CC/Discord/Concurrence Vs $\Delta$T with $T_1=0.2$ and $T_2=T_1+\Delta$T. $\Delta=0.3$, $\Gamma_1=\Gamma_2=0.05$, $\omega_1=\omega_2=1$.}
\label{fig4.1}
\end{figure}

In Fig \ref{fig4.1}, the temperature of one reservoir is kept fixed at 0.2 and the other at $0.2+\Delta T$. Comparison can be drawn with the equilibrium situation (Fig \ref{fig2}). In non-equilibrium case, the quantum discord and classical correlation no longer approach to zero as we increase the average temperature. This is noticed from the the solution \ref{sb} that the non-vanishing coherence term in energy basis $\lim_{T\rightarrow \infty}\rho_{23}=ig$ to the leading order will give the non-vanishing coherence term in the local basis. In the equilibrium situation, the coherence term vanishes completely in the high temperature limit. This suggests that the quantum nature of the system will remain as the temperature of one of the bath becomes high. The entanglement will still disappear at finite temperature bias. The rise and decay of the quantum correlations can be understand through the effective temperature defined in \ref{Teff}. As $\Delta$ T increases, the effective temperature to the leading order $T_{eff}=T_1+\Delta T/2$ increases, and we can refer the increase and decrease of the quantum correlations to the equilibrium case discussed in the last section. 

\subsubsection{The "averaged effect" of the nonequilibrium system}
We stated in the introduction that the nonequilibrium effects discussed in the previous studies are mainly due to the system being in equilibrium with the \textit{average} of two reservoirs. What do we mean by this statement? 

Physically, it means that the bipartite system is so strongly coupled within itself that the subsystems are always in equilibrium, and that the system reacts to the average of the two reservoirs as a whole. In comparison, when the inner-system coupling between the two subsystems is weak, the two subsystems can not be treated as in identical states, instead each subsystem is approximately in equilibrium with its own reservoir. The nonequilibrium effect arises from the two subsystems not being in an identical state. 

This can be noticed by the leading order solution given in Eqn.~\ref{bs}. All the population terms up to the leading order are functions of $n_{ip}$, which represents the \textit{sum} of the two reservoirs. When the tunneling rate is large, i.e. $g \ll 1$, the higher order terms and the coherence terms in Eqn.~\ref{bs} can be  ignored. Then the $0^{\text{th}}$ order population terms return to the equilibrium solution if we make the following substitution, 
\beq n_i(T)_{equilibrium} \rightarrow \dfrac{1}{2}n_{ip}=\dfrac{n_i(T_1)+n_i(T_2)}{2}, \label{tbeff} \eeq
where $n_i(T_j)$ is particle number of the $j^{\text{th}}$ reservoir at the energy level which equals to the $i^{\text{th}}$ eigenenergy of the system. This returns the equilibrium solution with two equivalent baths replaced by the average of two baths at different temperatures. Therefore, if we increase the temperature bias between the two reservoirs, the behavior of the quantum correlations shows instead of the nonequilibrium effect but the equilibrium effect with the average of the two reservoirs as it is essentially due to the same physics as in the equilibrium situation, which we give a full discussion in Appendix \ref{equilibrium}. 

The true nonequilibrium effect should result from the \textit{difference} between the two reservoirs instead of the sum/average. The \textit{only} term that depicts the bias of the two reservoirs is $n_{im}$, which appears in the \textit{coherence terms} that are ignored by the Lindblad solution and also the higher order terms that only become effective when $\Delta \not\ll \Gamma$, i.e. the coupling or the tunneling rate of the system is not too large for us to treat it acts as \textit{one whole system} weakly coupled with two reservoirs simultaneously. When the subsystems are not strongly coupled, then each subsystem is then approximately in equilibrium with its environment and the system will not behave the same way of the equilibrium scenario. As we will discuss later, to capture the essence of the nonequilibrium effect, we need to consider the case when $g \not\ll 1$ and set the environment at finite temperature to smooth out the resonance due to the chemical potential.

\begin{figure}[ht]
\centering
\subfloat[QD/QMI Vs T]{{\includegraphics[width=0.4 \textwidth]{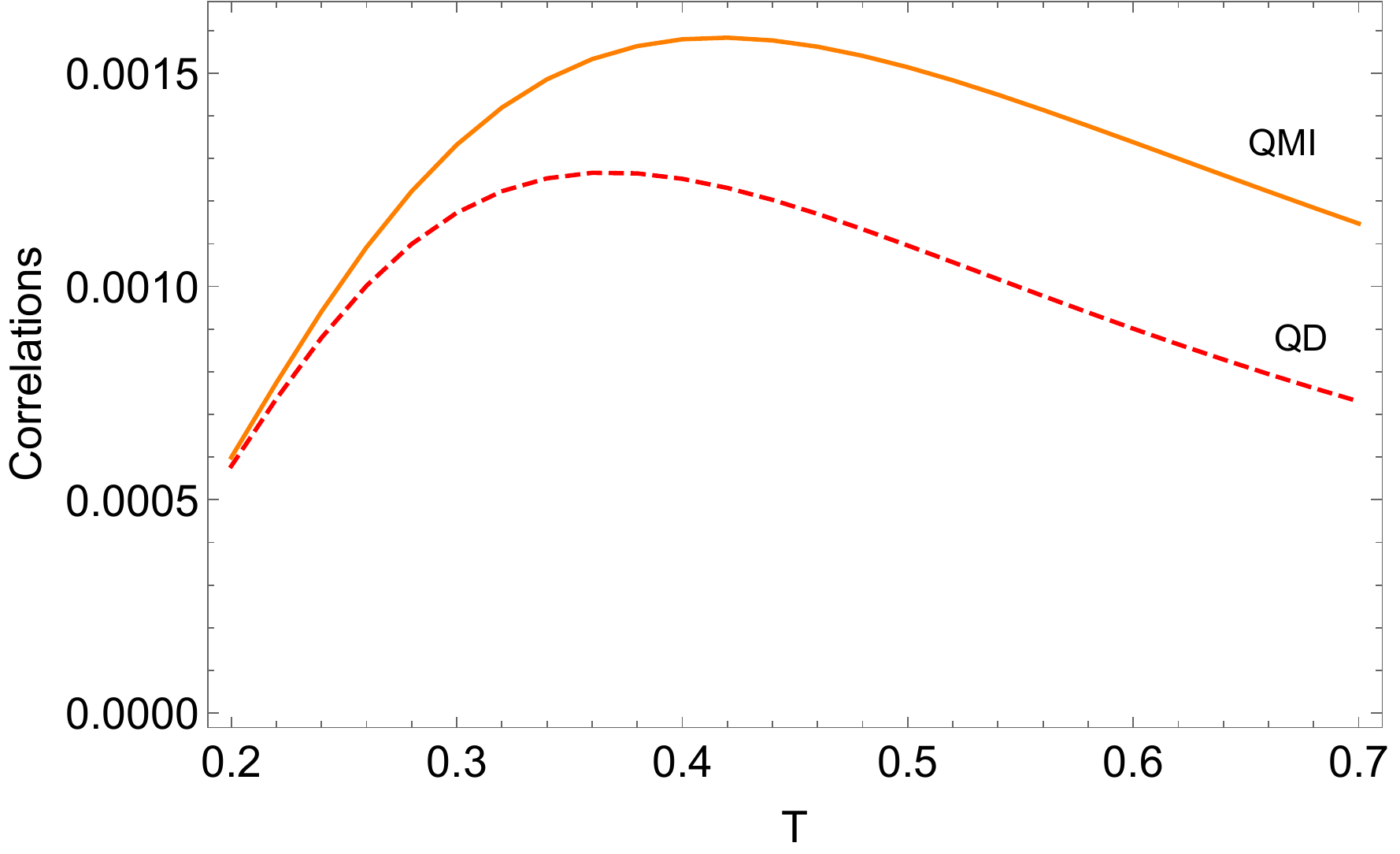}}}
\quad
\subfloat[QD/QMI Vs $\Delta$T at $T_1=0.5$]{{\includegraphics[width=0.4 \textwidth]{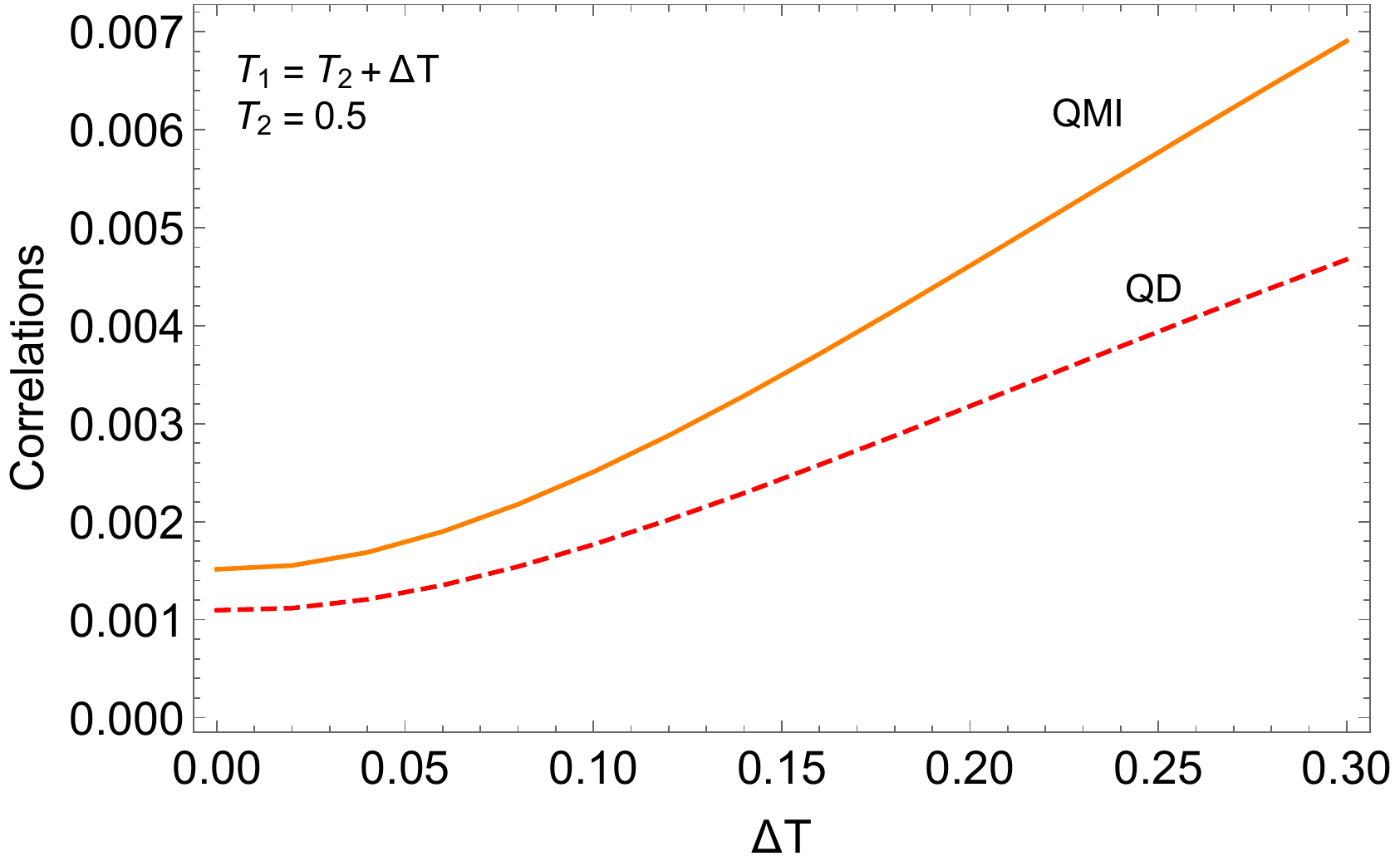}}}
\caption{(Color online) Separating nonequilibrium effect from the thermal effect. $\Delta=0.05$. (a) Discord and QMI Vs T in equilibrium. (b) Discord and QMI Vs $\Delta$T in nonequilibrium case. $T_2=0.5$ and $T_1=T_2+\Delta T$. Notice from (a) that the correlations decay after the average temperature exceeds 0.4, therefore \textit{turning up average temperature at T$>$0.4 suppresses quantum correlations} from the thermal excitation, however, correlations increase anyway.}
    \label{ne}
\end{figure}

\subsubsection{The "distilled" correlation generation due to nonequilibriumness}
When we relax the condition $g=\frac{\Gamma}{\omega_1'-\omega_2'} \ll 1$ and instead set the tunneling rate very small, e.g. $\Delta=\Gamma=0.05$, the coherence term $\rho_{23} \propto g=\frac{\Gamma}{2\Delta}$ in the leading order is not anymore negligible. On the other hand, the energy gap between the two non-localized eigenstates ($2\Delta$) is small and the two states are almost equally populated. The consequence is, in this extreme the entanglement due to the energy eigenstate populations (of which the leading term is $n_1+n_2$) can be ignored and all the entanglement generation are from the nonequilibrium effect. (In local basis, the coherence is $-\frac{1}{2}(\rho_{22}-\rho_{33})+\Im(\rho_{23})$ as given in Appendix \ref{local}, where $\rho_{22}-\rho_{33}$ is the difference in population of the two non-local states.) \textit{We point out that the generation of quantum correlation no longer comes from the averaged thermal effect of the reservoirs as but from the energy flow due to the nonequilibriumness of the system.} This is a distinctive mark of the nonequilibrium effect that the quantum correlations tend to increase monotonically with nonequilibriumness. E.g. see Fig \ref{fig4}(b). 

In Fig \ref{fig4}(b), as an example we plot the discord against $\delta$T at small tunneling rate (QMI and CC have similar monotonic trend).  The thermal generation of quantum correlations, which is indicated by the left-end value, is negligible compared with the right-end value of the graph. The averaged thermal effect can be indicated from Fig \ref{fig4}(b). The change of quantum discord due to the average temperature is less than a factor of two and is negative when $T_{avg}>0.3$, however the large $\delta T$ value in Fig \ref{fig4}(c) is orders of magnitudes larger than the equilibrium value.  Thus, we can say with confident that correlation generation is no larger due to the thermal excitation due to the average of the two reservoirs but the quantumness of the system enhanced by the equilibriumness.

\begin{figure*}[htb]
\centering
\subfloat[QMI Vs $\Delta$T at different chemical potentials]{{\includegraphics[width=0.32 \textwidth]{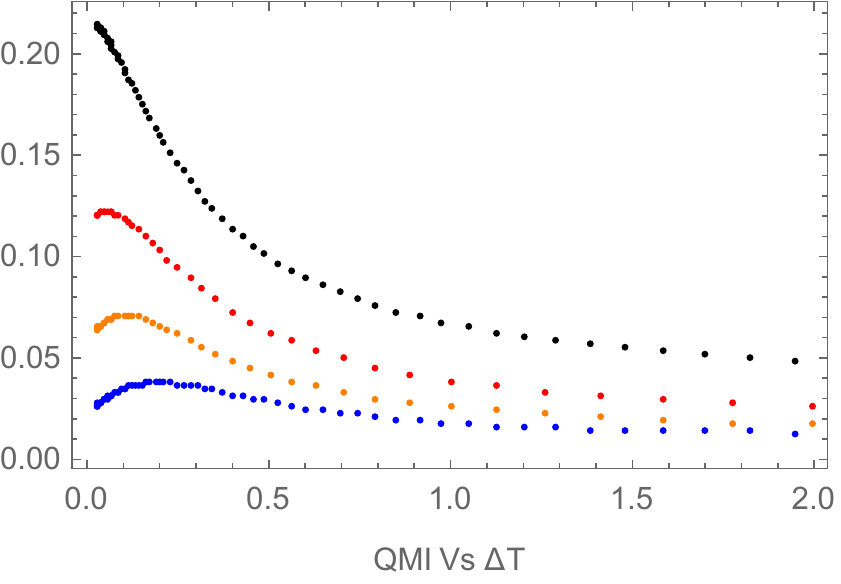}}}
\quad 
\subfloat[QMI Vs $\Delta$T at different $T_1$]{{\includegraphics[width=0.32 \textwidth]{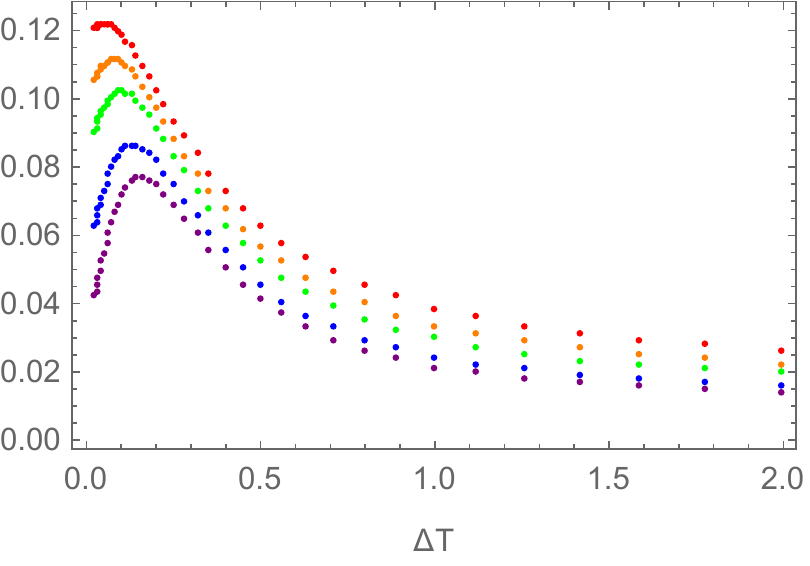}}}
\quad
\subfloat[CC/Discord/QMI/Concurrence Vs $\Delta$T]{{\includegraphics[width=0.32 \textwidth]{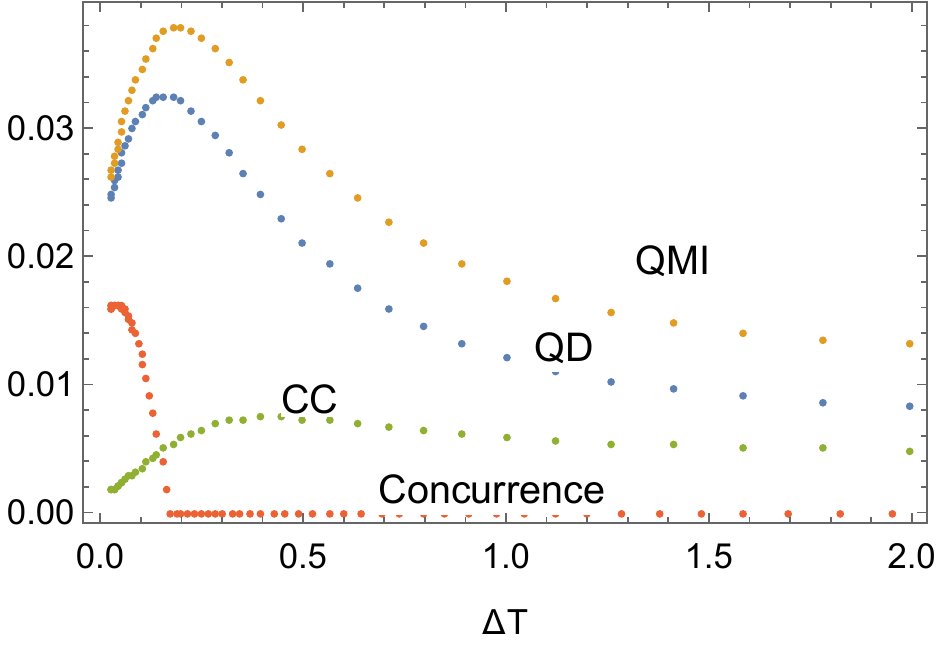}}}
\caption{(Color online) Nonequilibrium fermionic reservoirs. (a) QMI Vs $\Delta$T. From bottom to top, $\mu_1=\mu_2$=0, 0.2, 0.35, 0.5. The temperature T$_1$ is fixed at 0.2. (b) The "rainbow" diagram of QMI Vs $\Delta$T with $T_1$ set to 0.1, 0.12, 0.15, 0.17, 0.2, 0.3 from bottom to top. Here $\mu_1=\mu_2=0.35.$ (c) Concurrence(red), CC(green), discord(blue), and QMI(orange) from bottom to top are plotted as a function of $\Delta$T at $T_1=0.2$ and $\mu_1=\mu_2=0$. For all, the parameters are set to $\Delta=0.3$, $\Gamma_1=\Gamma_2=0.05$, $\omega_1=\omega_2=1$.}
    \label{fig5}
\end{figure*}

In Fig \ref{ne}, we separate the quantum correlations due to thermal effect, Fig \ref{ne}(a) and correlations due to nonequilibriumness Fig \ref{ne}(b). As we notice, when the temperature is larger than 0.4, QMI and QD decrease with the increase of T. If we pick T=0.5, from Fig \ref{ne}(a) the quantum correlations due to thermal effect decrease with temperature. Now we pick $T_2=0.5$ and $T_2=T_1+\Delta T$, according to \ref{Teff} the effective temperature increases, and the effect from the thermal excitation will result in the weakening of quantum correlation as $\Delta T$ increases. However, them become enhanced as shown in Fig \ref{ne}(b). Therefore, the increase of the generation of quantum correlations can not be from the same sources as in the equilibrium case. This is the intrinsic nonequilibrium phenomenon. The temperature bias enables a nonzero quantum flux \textit{inside the system} from site attached to the high temperature reservoir to site attached to the low temperature reservoir. The nonzero flux between the sites, instead of the thermal excitation, contribute to the coherence of the system.

\section{Quantum correlations of fermionic system in nonequilibrium environments (Fermionic reservoirs)}
For fermionic reservoirs, all the arguments given in the previous bosonic case still hold. The solution of the steady state density matrix for the system in fermionic baths up to the leading order in $g$ is given as follows,
\beq \bsplit  &\rho_{11}= (1-n_{1p}/2) (1-n_{2p}/2)+\mathcal{O}(g^2)\\
&\rho_{22}=(n_{1p}/2 - (n_{1p} n_{2p})/4)+\mathcal{O}(g^2)\\
&\rho_{33}=(n_{2p}/2 - (n_{1p} n_{2p})/4)+\mathcal{O}(g^2)\\
&\rho_{44}=(n_{1p} n_{2p})/4+\mathcal{O}(g^2)\\
&\rho_{23}=-i(n_{1m} + n_{2m}) g/2+\mathcal{O}(g^2), \end{split} \label{fs}\eeq
 $n_{i,p/m}=n(\omega'_i,T_1,\mu_1)\pm n(\omega'_i,T_2,\mu_2)$ and $n(\omega'_i,T_j,\mu_j)$ follows Fermi-Dirac distribution. The bosonic bath solution has limited range of applicability due to the non-compactness of occupation number $n_i(T)$. The higher order terms involves cubit terms of $n_i(T)$ and thus their values can easily surpass the leading order. The fermionic occupation has a compact range and the solution in general gives a better approximation (later we will see that it gives the qualitatively correct behavior if $g<\frac{1}{4}$).

The fermionic reservoirs solution, like the bosonic case, can also return to the equilibrium solution with $n_i$ replaced by $n_{i,p}$ i.e. we can identify $n(\omega'_i,T,\mu)_{equilibrium} \rightarrow n_{i,p}/2=(n(\omega'_i,T_1,\mu_1)+ n(\omega'_i,T_2,\mu_2))/2$. Physically it means that the effect of the two equivalent reservoirs are now replaced by the average of the two different reservoirs.  

This section is arranged as follows. In part A, we discuss the quantum correlations at large tunneling rate and comment on the transition of quantum entanglement at large chemical potential biases. In part B, we discuss the breakdown of Lindbladian assumption in small tunneling rate case and show the distinguishable traits of quantum correlations with the large and small tunneling. In part C, we discuss the transition from the large to the small tunneling behaviors and the boundary between the two cases. Besides, we show that near the boundary the entanglement may resurrect due to the nonequilibrium effect after its decay to zero from the averaged effect, and briefly comment on the energy current perspective and extremely low temperature scenarios.

\subsection{Large tunneling rate (average effect)}
\subsubsection{The averaged effect on quantum correlations}
If we assume the tunneling rate is much larger than the decay rate, then Large tunneling rate results in the large splitting between the two non-local eigenstates. As a result, the coherence $\rho_{23}$ in the energy representation is small, see solution \ref{fs}. In this case, we can approximate the two reservoirs effect at a given temperature and chemical potential with one reservoir with temperature $T_{eff}$ and chemical potential $\mu_{eff}$. At zero or very low chemical potential, the quantum correlations behave similarly to that of the bosonic environment as explained in the equilibrium section, see Fig \ref{fig5}. As the chemical potential increases, the system deviates from the bosonic bath case. Chemical potential describes the tendency of the reservoirs to give or admit an electron. Since the reservoirs can give electrons to the system and accept electrons from it, tuning chemical potential can give arise to the more efficient particle inputs or outputs to or from the system. This can be seen by comparing the magnitude of the correlations in Fig \ref{fig4}(a) and Fig \ref{fig5}. As the chemical potential increases, the correlations reach their maxima at lower and lower temperatures until zero. We witness such change of monotonicity with the increasing of chemical potential, see Fig \ref{fig5}(a,b).

When raising the temperature of one of the reservoirs, we effectively increase the average effective temperature of the system. The fermion occupation number at the site in contact with the hotter bath increases and through the inter-site tunneling, the system gets more correlated. As the chemical potential becomes higher, the temperature tends to erase the already optimized quantum correlations. The maximal point shifts to lower and lower temperatures until it reaches zero, Fig \ref{fig5}(a). When the effective temperature becomes higher, energy gap that distinguishes a local state from the non-local state is washed out. The state becomes more localized and quantum correlations decay, see Fig \ref{fig5}(a,b,c). Fig \ref{fig5}(c) shows the non-monotonic behavior of the four correlation measures. Quantum entanglement vanishes at finite temperature bias, the other three correlations asymptotically reach a nonzero value due to the finite coherence caused by the current of particles. The nonzero asymptotic value of correlations was explained in the bosonic reservoir case, for chemical potential bias (see Fig \ref{sm1}(a)), it creates an imbalance in the occupation numbers on the two ends, and also contribute to the coherence \ref{fs}. 

\begin{figure}[htb]
\centering
\subfloat[Entanglement Vs $\Delta \mu$]{{\includegraphics[width=0.4 \textwidth]{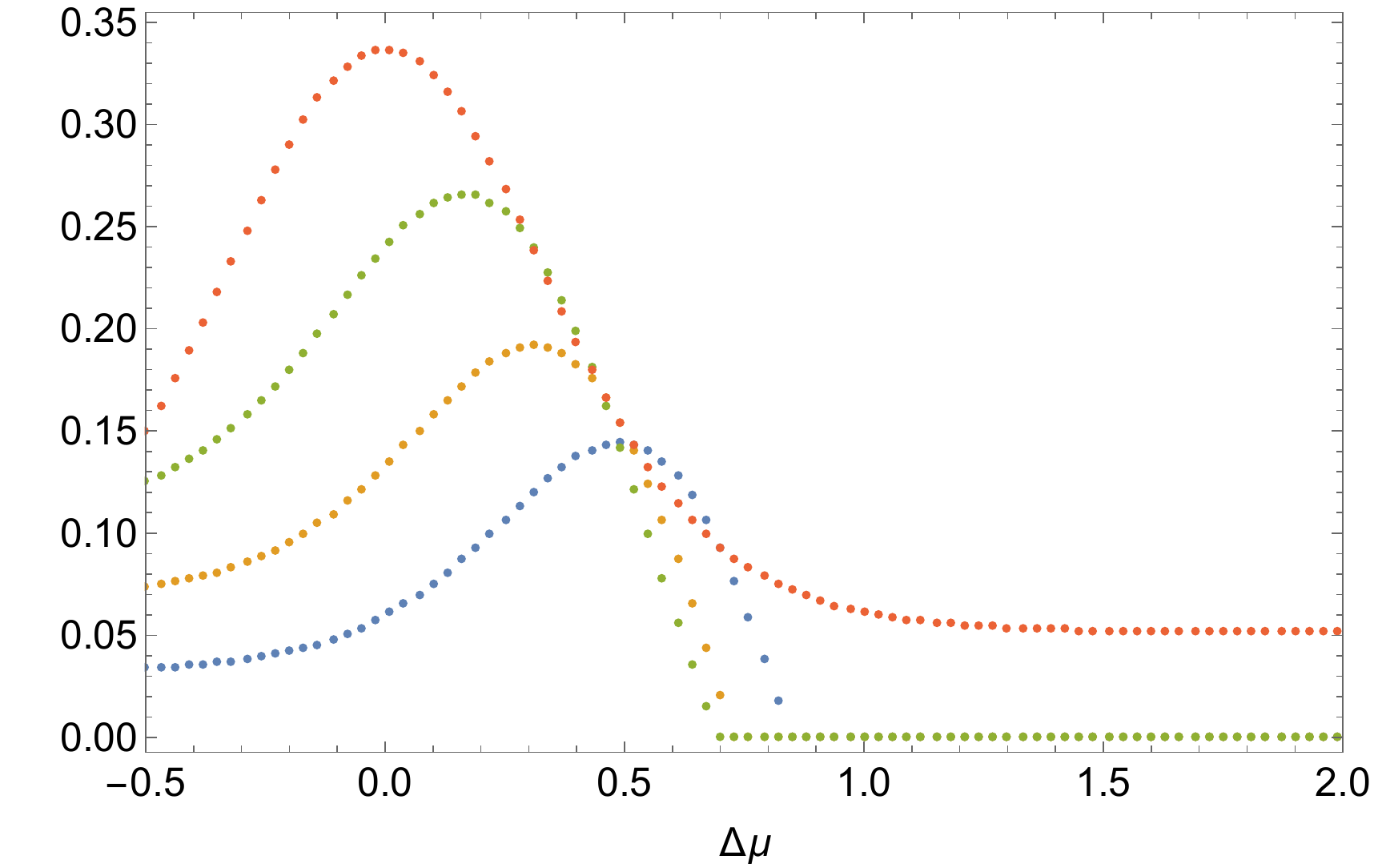}}}
\quad \quad
\subfloat[Transition to entanglement at asymptotic value $\Delta\mu=\infty$ Vs $ \mu_1$]{{\includegraphics[width=0.41 \textwidth]{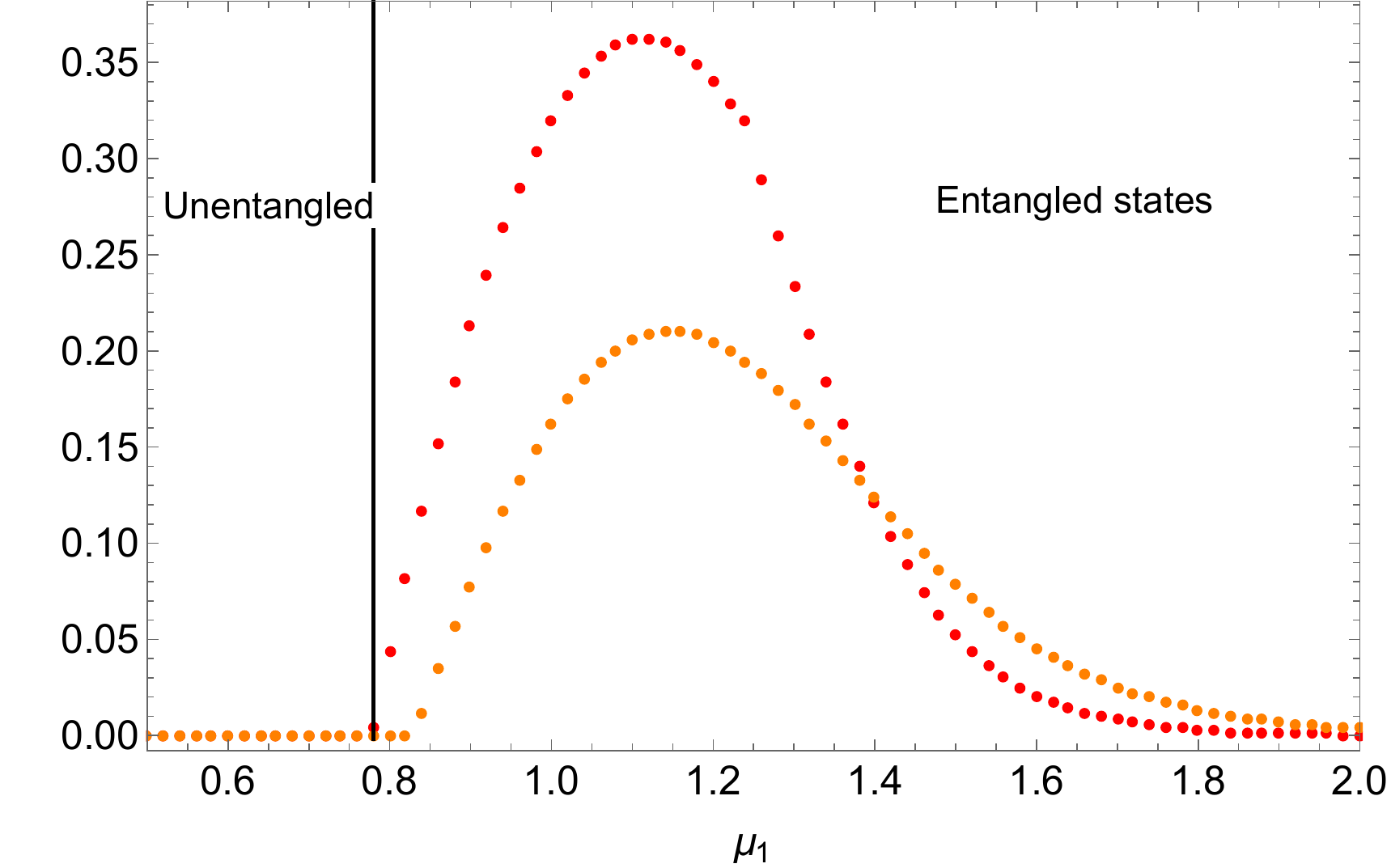}}}
\caption{(Color online) Transition of quantum entanglement. (a) Concurrence Vs $\Delta \mu$ at $\mu_1=$0.3, 0.5, 0.7, 1 from bottom to top.  $T_1=T_2=0.2$. The concurrence will not stay finite when $\mu_1$ is lower than a critical value. (b) The concurrence at $\Delta \mu=\infty$ with $\mu_1$. From left to right, the temperature is set to $T_1=T_2=$0.1, 0.15. For both graphs, the parameters are set to $\Delta=0.3$, $\Gamma_1=\Gamma_2=0.05$, $\omega_1=\omega_2=1$.}
    \label{ct}
\end{figure}

When $\Delta/\Gamma \gg 1$, we can refer to the approximate solution \ref{fs}. When the temperatures of the reservoirs are fixed, a redefinition $n(\omega'_i,T_{i,eff},\mu_{i,eff})\equiv n_{i,p}/2=(n(\omega'_i,T_1,\mu_1)+ n(\omega'_i,T_2,\mu_2))/2$ will return the solution back to the equilibrium case, except for the coherence term which is of higher order in the expansion of $g=\frac{\Gamma}{\omega_1'-\omega_2'}$. With chemical potential of one of the reservoirs fixed, increasing chemical potential bias equivalently raises effective average chemical potential of the two reservoirs. This explains why increasing chemical potential bias has the similar effect to the system as increasing the chemical potential in the equilibrium case, Fig \ref{sm1}(a) and Fig \ref{ct}(a). When the reservoirs have very high chemical potential the sites are almost fully occupied, the tunneling of fermions stops and quantum correlations are degraded.

\subsubsection{Entanglement not always dies}
On the other hand the entanglement can still vanish to zero in an abrupt manner if we tune the chemical potential of one of the reservoirs, though in equilibrium case it decays in a smooth exponential way. As shown Fig \ref{ct}(a), when $\mu_1$ is small, concurrence dies at finite bias. However a transition is noticed that when the chemical potential $\mu_1$ reaches some critical value, the entanglement will no longer witness the sudden disappearance if $\mu_1$ is larger than the value. In fact, it will remain finite no matter how we tune the chemical potential of the second reservoir. This discontinuity in the asymptotic behavior of concurrence can be seen in the leading order in the expansion of $\Gamma/\Delta$. The critical chemical potential above which concurrence will remain finite can be calculated by requiring 
\beq \lim_{\mu_2 \rightarrow \infty} \mathcal{E}(\beta_1,\beta_2,\mu_1,\mu_2)>0, \eeq
where $\beta_i=1/T_i$. For the case $\omega_1=\omega_2=\omega$, this gives us
\beq \mu^*=\omega+\Delta-T_1 \ln{(\sqrt{2e^{4 \beta_1 \Delta}-4e^{2 \beta_1 \Delta}+2}-e^{2\beta_1 \Delta}-1)}, \label{critmu}\eeq
where $\omega$ is the energy of the free excitation of one fermion and $T_1$ is the temperature of the reservoirs with a finite chemical potential.

\begin{figure*}[htb]
\centering
\includegraphics[width=0.33 \textwidth]{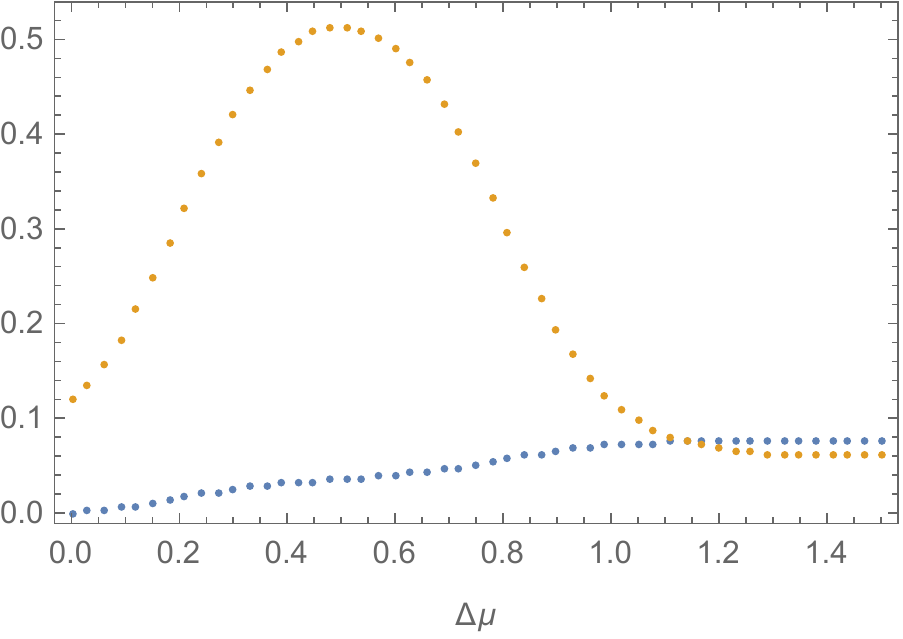}
\quad \quad
\includegraphics[width=0.33 \textwidth]{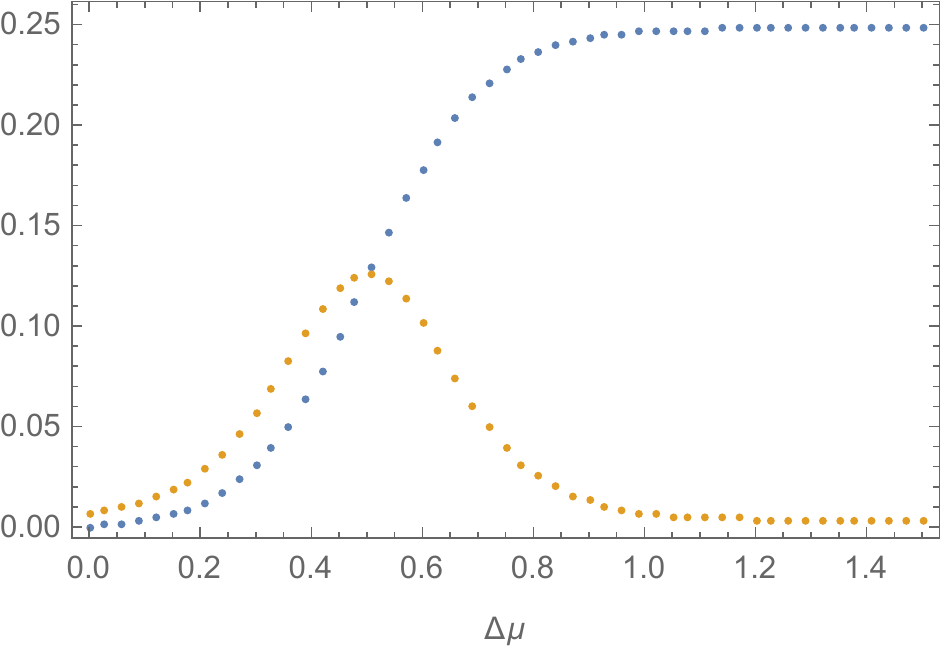}
\caption{(Color online) Contributions to the local site coherence from the population term $\rho_{22}-\rho_{33}$(orange) and coherence $\rho_{23}$(blue) at (a) large tunneling rate $\Delta=0.3$ and (b) small tunneling rate $\Delta=0.05$. For both graphs, $\mu_1=$0.5, $\mu_2=\Delta\mu+\mu_1$ and $T_1=T_2=0.1$.}
    \label{coherence}
\end{figure*}

We can check that $\mu^*$ monotonically increases with temperature $T_1$. The minimal transition potential is when the temperature $T_1=0$, this gives us \beq \mu^*_{min}=\omega-\Delta, \label{ctmin}\eeq which corresponds to the lowest energy of the non-local state. Notice that $\mu^*$ only has real-valued solution when $T_1$ is lower than the critical temperature defined as, 
\beq T_{critical}=\dfrac{\Delta}{\ln (\sqrt2+1)}. \eeq
This critical temperature is also where the concurrence vanishes in equilibrium situation as given in the previous section \ref{criticalT}.  The critical chemical potential is only dependent on the temperature of the first bath and not on the second bath which has the higher chemical potential. In Fig \ref{ct}(b), we plot the asymptotic behavior of the concurrence with the increase of chemical potential of one reservoir at different temperatures, where the line dividing the entangled and unentangled state is for the $T=0.1$. When $\mu_1$ is larger than the $\omega'_2$, though the system is still entangled, the entanglement rapidly decreases, and for $\mu_1 \gg \omega'_2$, the entanglement is essentially zero. 

We should notice that the above result are only valid when $\Delta\gtrsim \Gamma$. For the parameters under consideration, it gives a very accurate approximation. But we will see in the following subsection that when $\Delta \lesssim \Gamma$, all the conclusions drawn in this subsection is not anymore valid. In fact, we will see in the next subsection that all the quantum correlations take a different trend when the tunneling is sufficiently small.

\begin{figure*}[ht]
\centering
\subfloat[QD/QMI Vs $\Delta \mu$ at large tunneling rate]{{\includegraphics[width=0.33 \textwidth]{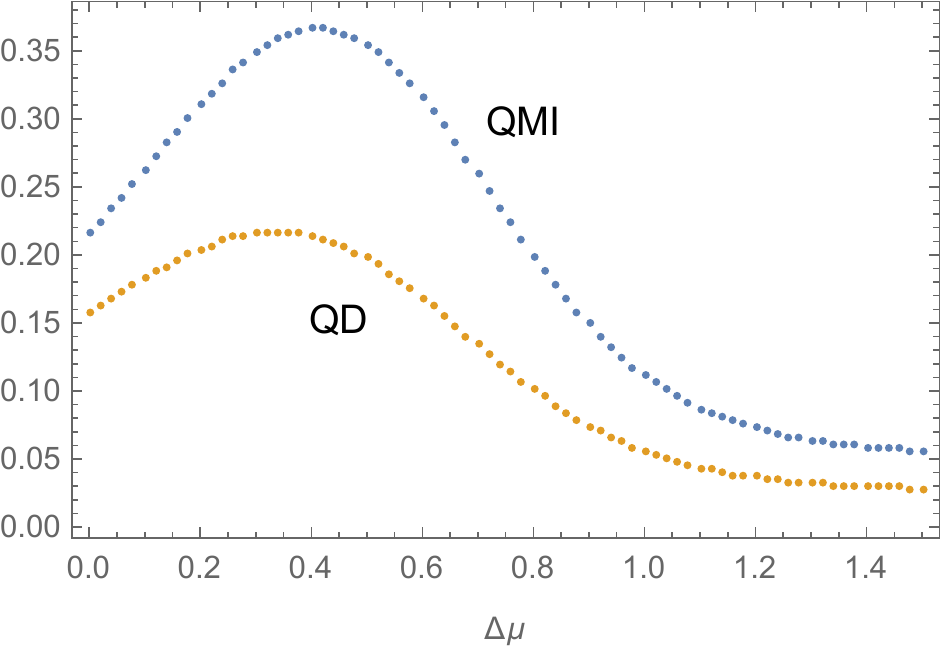}}}
\quad 
\subfloat[QD/QMI Vs $\Delta \mu$ at small tunneling rate]{{\includegraphics[width=0.34 \textwidth]{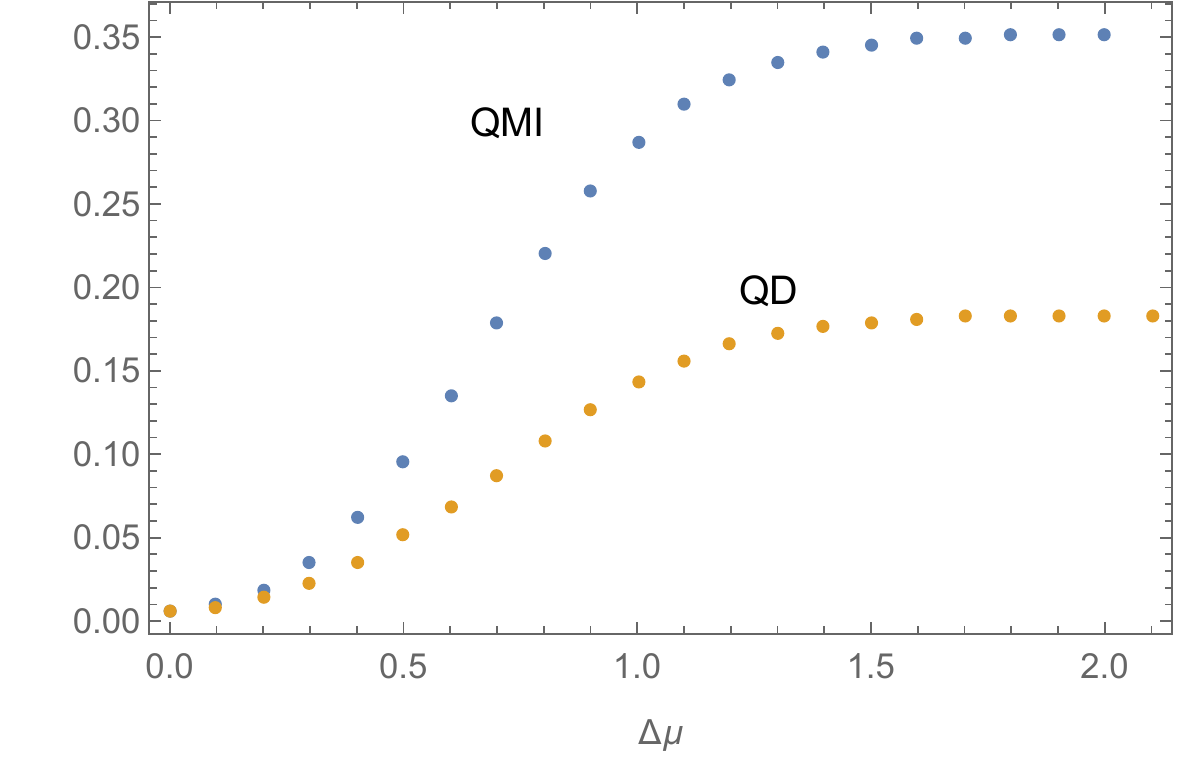}}}
\quad
\subfloat[Concurrence Vs $\Delta \mu$ at small tunneling rate at two temperatures]{{\includegraphics[width=0.33 \textwidth]{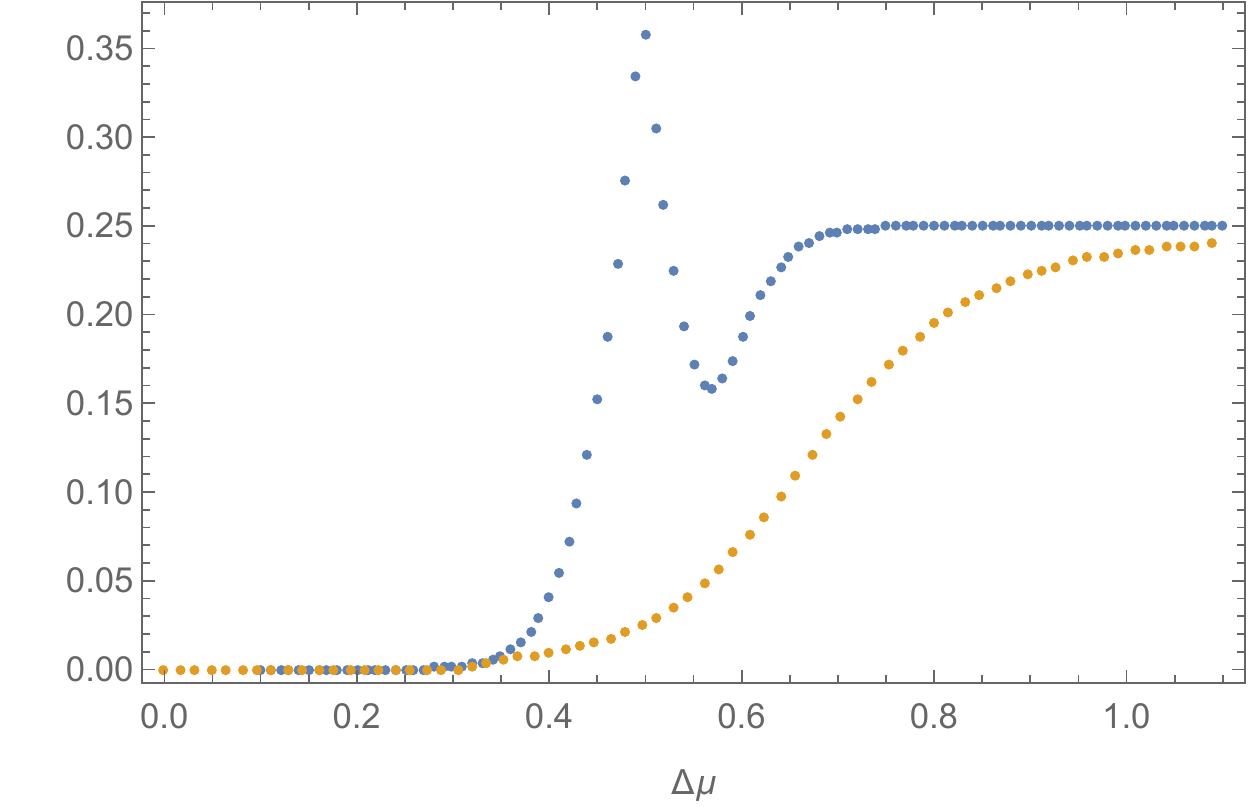}}}
\quad
\subfloat[Concurrence Vs $\Delta \mu$ at large tunneling rate at two temperatures]{{\includegraphics[width=0.33 \textwidth]{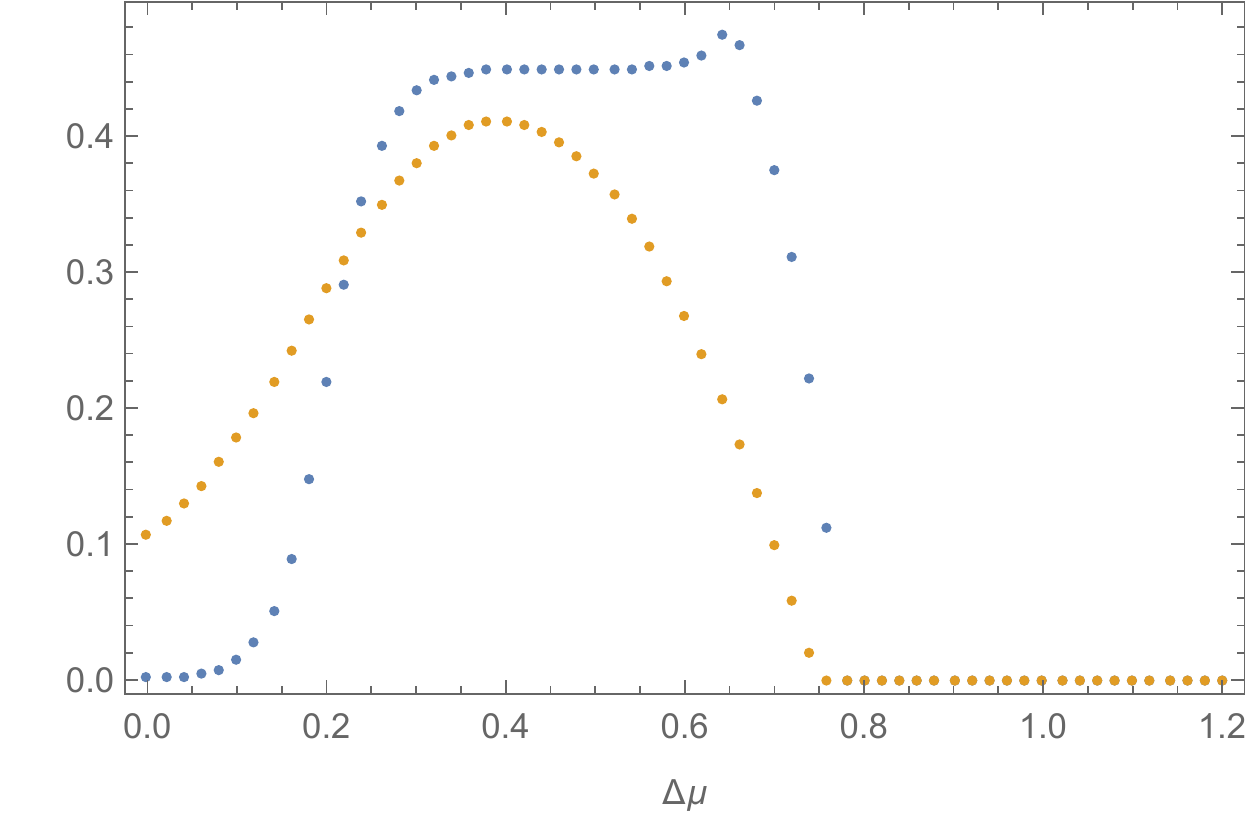}}}
\caption{(Color online) \textit{Separating nonequilibrium effect and averaged effect.} (a)  QMI (blue, upper) and QD (orange, lower) Vs $\Delta \mu$ at $\Delta$=0.3. (b)QMI (blue, upper) and QD (orange, lower) Vs $\Delta \mu$ at $\Delta=0.05$.  For (a) and (b), $T_1=T_2=0.2$, $\mu_1=0.5$. (c) Entanglement Vs $\Delta \mu$ at $\Delta$=0.05. (d) Entanglement Vs $\Delta \mu$ at $\Delta=0.3$. For (c) and (d), $T_1=T_2=0.03$ is the blue line (with sharp turnings), $T_1=T_2=0.1$  is the orange (smooth line), and $\mu_1=0.5$. \textit{Lowering the tunneling rate can in fact increase the correlations at large $\Delta \mu$.}}
\label{sm1}
\end{figure*}

\subsection{Separating the averaged and nonequilibrium generation of correlations}
The nonzero quantum correlations at small tunneling rate and at large tunneling rate are due to completely different mechanisms. The physical picture is the following, when the tunneling rate (or coupling) of the system is large, the typical time scale for the system interaction is much shorter than that of the system-environment coupling. Thus, the two subsystem is approximately in equilibrium, and the system as a whole acts to the average of the two reservoirs. When the tunneling rate of system is smaller than the system-environment coupling, then each subsystem is approximately in equilibrium with its reservoir. The two subsystems can not be treated as in an identical state interacting as a whole. In this scenario we see not only the average of the two environments but also the difference between them coming into play.

\subsubsection{Insufficiency of Lindbladian} \label{spcg}
In the previous sections, we give an explanation of "distilled" quantum correlations at small $\Delta$ based on the reduced density matrix and the basis transformation given in Appendix \ref{local}. Here we give a detailed explanation based on the arguments given in the beginning of this session. Let $\tau_{S}\approx \frac{1}{\omega_1\prime-\omega_2\prime}$ be the typical time scale that characterizes one fermion traveling between site S1 and S2. When the tunneling rate is small, i.e. $\Delta < \Gamma_i$, $\tau_S$ is longer than the time of the system-reservoir interaction $\tau_{SR}\approx \frac{1}{\Gamma}$. Intuitively, it means that the communication within the system is slower than that between the reservoirs and system. \textit{This means that the secular approximation which require the system interaction time to be much shorter than the relaxation time is no longer valid}. Though the Lindablad equation guarantees the dynamical semigroup structure, the positivity is not an issue in many situations. On the other hand it sacrifices the important effects of the nonequilibrium physics and becomes very inaccurate in this situation. The breakdown of the assumption for secular approximation is equivalent to the existence of a strong nonequilibrium condition inside the system. In this case, each site is approximately in equilibrium with its reservoir but the two sites can not be treated as one tightly bounded system. Consider the case when the average thermal generation of the correlation is very low, when the chemical potential bias increases, the two sites start witness nonzero flux (see Fig\ref{crnt} in Appendix D) and correlation increases. This generation is the result of nonequilibriumness, as the magnitude of correlation depends on the difference of the two reservoirs instead of the sum. This flux easily saturate as $\Delta \mu$ increases further due to the small tunneling rate, and the correlation remains at a constant value Fig\ref{sm1}(b). The discussion of energy current is provided in Appendix \ref{current}.

With the above argument, we would naively expect stronger correlation within the system with the increase of chemical potential bias, but this is not always true from the previous discussions. When the tunneling rate of the system is very large, i.e. $\Delta \gg \Gamma_i$. The tunneling time of the system is negligible, i.e. $\tau_{S}\approx \frac{1}{\omega_1\prime-\omega_2\prime}\ll \tau_{SR}$. This is identical to the secular approximation, where we ignore the non-secular coherence terms and also the influence brought by the system's not being in equilibrium. In this case, the two fermion sites can effectively be treated as one system coupled with the two reservoirs simultaneously, and the population on each site is determined by the average of the two bath. In this case, the sites can no longer be viewed as in approximate equilibrium with their respective reservoirs but in equilibrium within the two sites. Therefore, when one of the bath reaches the resonant energy of the system, the quantum correlations are boosted. With the parameter used in Fig \ref{sm1}(a,d), this corresponds to the range from 0.2 to 0.8. If we get ride of the smooth-out effect of finite temperature, it is more obvious (blue line in Fig \ref{sm1}(d)). The large tunneling rate kills the non-secular terms and nonequilibrium effect, the correlations die after the resonance.

Finally, we remark that by setting tunneling rate low, we separate out the usually-ignored nonequilibrium terms as they become the dominant contribution in the large bias situation. Furthermore, we notice that the remaining averaged resonance effect can be erased by tuning up temperature (noted in Fig\ref{fig2}(e,f)), we distill the correlation (entanglement) generation and obtain the contribution purely from the nonequilibrium effect (orange line in Fig\ref{sm1}(c)).   This procedure avoids the averaged thermal effect of the two baths and exhibits the correlation generation from only the nonequilibriumness-related terms.

\subsubsection{Numericals and plots}
The small tunneling rate picture gives a zoomed-out view of the large tunneling scenario, see Fig \ref{sm1}(c,d). The resonance peak at a larger tunneling rate (Fig \ref{sm1}(d)) corresponds to the peak near the resonance of Fig \ref{sm1}(c). It is not a simple zooming-in of the spiky peak (blue line) but overshadows the nonequilibrium effect of the monotonic increase trend with bias. The resonance effect, blue lines, can be washed out by adding temperature to the reservoirs, orange lines. At a finite temperature, the resonance peak is erased, the leftover correlation generation only comes from the nonequilibrium effect for the small tunneling scenario and is monotonic with bias. While a large tunneling mimics the equilibrium case with a bell-like curve with effective chemical potential $\mu_{eff}$ increased. The peak, whose range is more obvious at low temperatures, corresponds to the resonance energy range at Site 2, i.e. $\mu_2\in [1-\Delta,1+\Delta]$. In this case, it is $\Delta \mu \in [0.2,0.8]$ which can be compared with the numerical results Fig\ref{sm1}(a,d). 

Another feature to notice is that, for large coupling Fig\ref{sm1}(a), the initial value which corresponds to equilibrium case is nonzero. For the small tunneling case Fig\ref{sm1}(b), the initial value is zero, i.e. no correlation from the was generated from the equilibrium effect. This correlation at later times is the result of the nonequilibriumness in contrast with Fig\ref{sm1}(ab). As we increase the temperature, all the sharp turnings are smoothed out, and we have a smooth monotonically increasing curve at small tunneling and a bell-like curve at a larger tunneling that appeared in the last subsection. We plot the entanglement at low temperature, for QD and QMI, the behavior is the same. In contrast to the previous results \cite{wuwei1}, we find from comparing the values of Fig\ref{sm1}(a) and (b), that \textit{lowering the tunneling rate between the two sites can in fact increase their correlations at large $\Delta \mu$}. Intuitively, we would expect that a large tunneling rate give a stronger quantum correlation. It is a most dramatic exhibition of nonequilirbium effect. 

\subsection{From large to small tunneling rates transition}
\subsubsection{Transition from small to large tunneling rates}
We discussed the distinctive behaviors for large and small tunneling rates for the last few sections. For the small tunneling, quantum correlations monotonically increase with the chemical potential bias, except for a local peak boosted by the resonance that is annihilated by temperature. For the large tunneling rate, QMI, QD, CC decay monotonically after the peak, and the entanglement dies at finite potential bias. Where is the \textit{boundary} between the two limits? 

We can estimate it by the following argument. At small tunneling rate, the quantum correlations increase monotonically with nonequilibriumness instead of following the transition rule given in \ref{ctmin}. We require that the entanglement at infinite chemical potential bias is nonzero even for arbitrarily small $\mu_1$, i.e.
\beq \mathcal{E}(T_1\rightarrow 0,T_2,\mu_1=\epsilon,\mu_2=\infty)>0, \eeq 
for arbitrarily small $\epsilon$ and finite $T_2$. Use the exact solutions given in Appendix \ref{sf}, and we arrive at the critical tunneling rate
\beq \Delta^*=2\Gamma. \eeq
For $\Gamma=0.05$, $\Delta^*=0.1$. Roughly speaking, when $\Delta$ is larger than $2\Gamma$, the quantum correlations are dominated by the resonance behavior. When $\Delta<2\Gamma$, the quantum correlations are more dominated by the coherence caused by the nonequilibriumness and increase as the nonequilibriumness enhances. What about when $\Delta \approx 2\Gamma$?

\begin{figure}[h]
\centering
\subfloat[Resurrection of entanglement near the critical tunneling rate.]{\includegraphics[width=0.39 \textwidth]{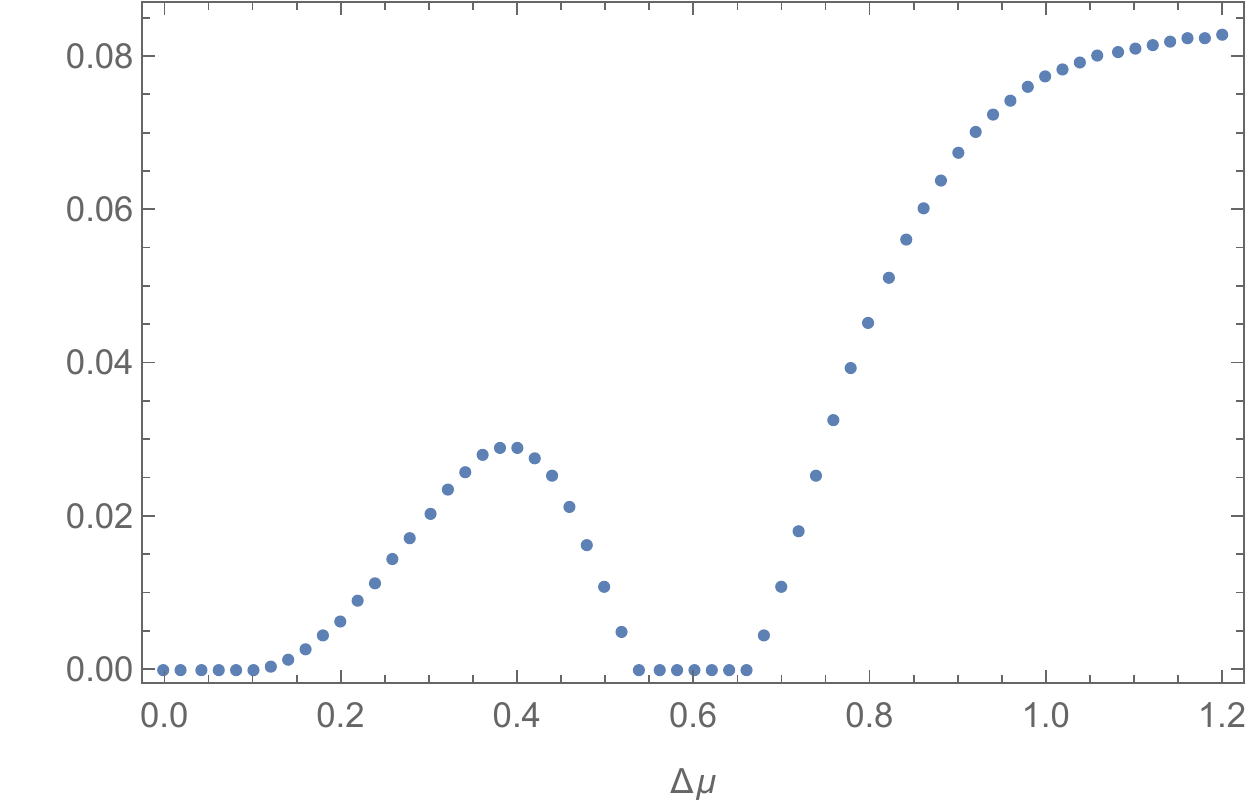}}
\quad
\subfloat[Non-smooth generation of entanglement at finite $\Delta \mu$]{\includegraphics[width=0.39 \textwidth]{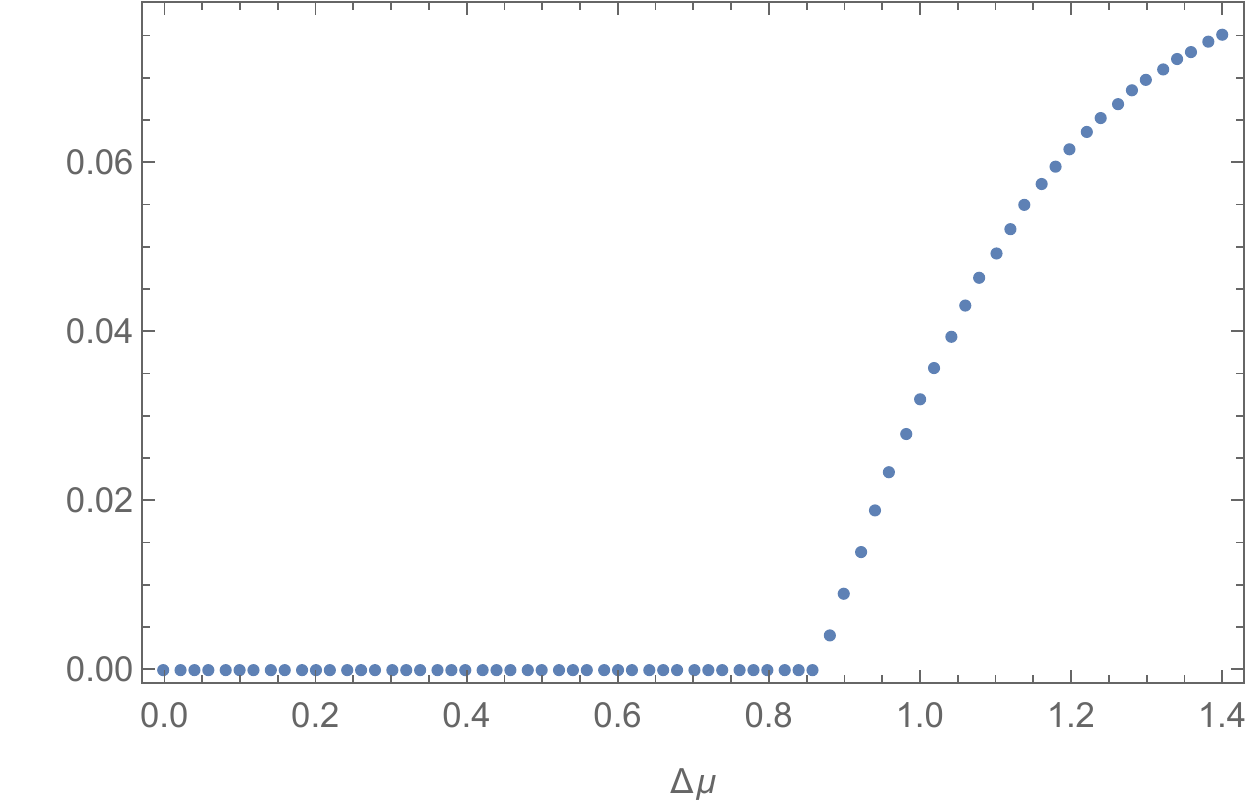}}
\caption{Resurrection and appearance of entanglement at a finite $\Delta \mu$. (a), $T_1=T_2=0.1$. (b), $T_1=0.1,\ T_2=0.2$. For both, $\Delta$=0.08, $\Gamma_i$=0.05, $\mu_1=0.5$ and $\mu_2=\mu_1+\Delta \mu$.}
\label{sm3}
\end{figure}

When $\Delta \ge 2\Gamma$, the entanglement is strictly zero after a finite nonequilibriumness is achieved (for example when chemical or temperature bias reaches some certain value). When $\Delta \lesssim 2\Gamma$ however, the entanglement dies at finite chemical potential bias, but it resurrects later a small period as the bias keeps increasing. After that, it is dominated by the nonequilibrium coherence, and monotonically increases to an asymptotic value, see Fig \ref{sm3}(a). The two peaks are however the results of two completely different sources. The first peak is the peak that appear in all large tunneling and equilibrium scenarios, which arises due to the resonant chemical potential reached by the second reservoir. The second peak is due to the nonequilibrium enhancement of the correlation. This plot exhibits the mixture of both effects. 

\begin{figure}[ht]
\centering
\includegraphics[width=0.4 \textwidth]{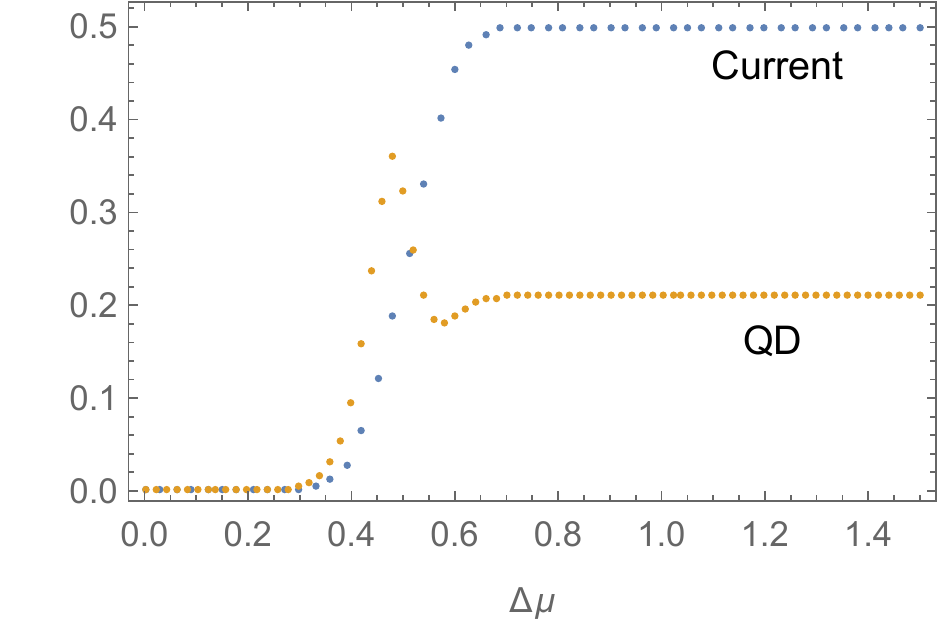}
\quad 
\includegraphics[width=0.4 \textwidth]{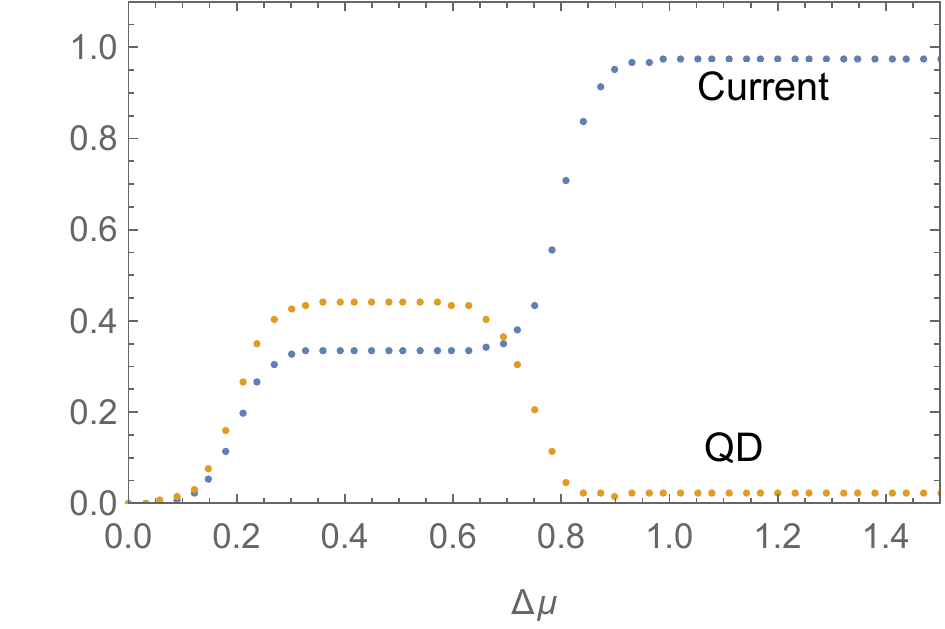}
\caption{(Color online) Discord (orange) and energy current (blue, upper at the right end) at large and small tunneling rates. (a) $\Delta=0.05$. (b) $\Delta=0.3$. For both plots, $T_1=T_2=0.03$ and $\mu_1=0.5$.}
\label{crnt2}
\end{figure}

As we have shown, raising the temperature of (one of) the reservoir(s) kills resonance peak, see Fig \ref{fig2}(e,f). Similar effect appears again in the graph Fig\ref{sm3}(b), and the entanglement is only seen as the chemical potential bias reaches some certain threshold value. As to the quantum discord and mutual information, they do not decay to exact zero with the increase of chemical potential bias after the first peak as the quantum entanglement does, thus, it does not have the dramatic qualitative changes in the intermediate tunneling regime. For discord and mutual information, this intermediate tunneling rate region roughly corresponds to the boundary whether the correlations at an arbitrarily low temperature will rise to some noticeable value after the post-resonance dip or decay monotonically. Such change of behaviors can be seen from Fig\ref{crnt2}.

\subsubsection{Energy current perspective and extremely low temperature situation}
Temperature smooths out the energy spectrum of the particles, adds disorder to the system and kills entanglement at finite temperature. At very low temperature, energy current provides another perspective to understand the change of quantum correlations. At low temperature, the system can only accept a particle when the chemical potential of the system reaches the energy of an eigenstate, and the current is not anymore smooth, see Fig \ref{crnt2}. At large tunneling $\Delta$, e.g. when $\Delta=0.3$, the two corresponding energy eigenvalues are 0.7 and 1.3. The energy current jumps to higher values when $\mu_2$ reaches the corresponding energies, and stays constant at other places. The plateau of the discord corresponds to when the chemical potential of the second reservoir reaches the energy of the first non-local state. The step-down of quantum discord corresponds to where the chemical potential is greater than the energies of both non-local eigenstates. The system is approximately in fully occupied state. Due to the large energy gap, there is little coherence between the two energy levels, and quantum discord decays dramatically after the resonance range. As shown in Fig \ref{crnt2} (b), the quantum discord at large $\Delta \mu$ is nonzero but negligibly small. In comparison, when $\Delta$ is small, the quantum discord still increases after the resonance range and its asymptotic value is not anymore negligible. In fact, when we increase the temperature, the resonant peak will be washed out (see Fig \ref{sm1}), and the coherence caused by the nonequilibriumness is the dominant source of quantum correlations of the system.

\section{Remarks and Conclusion}
In this paper, rich phenomena of quantum correlations at various tunneling rates, temperatures and chemical potentials of a two-fermion system are discussed. The nonequilibrium effect of the generation of quantum correlations is most apparent at small tunneling rates and it comes from the coherence of two eigenstates, while that at large tunneling rate is mainly due to the averaged effect of the two reservoirs. The transition between the two happens at $\Delta/\Gamma=2$, where quantum correlations may show a resurrection with the the breaking of equilibriumness.

To recapitulate, we explicitly showed that entanglement can be larger than quantum discord in an open system, which suggests that the interpretation of quantum discord as entanglement plus other quantum correlations is not faithful. In the equilibrium case, the quantum correlations of the system peak when the chemical potential approaches the frequency of the single-particle excitation of the system, otherwise they decay exponentially with the chemical potential. The quantum entanglement vanishes in a ``sudden death" manner with the increase of temperature. In the nonequilibrium situation, we separate the correlation generation due to the nonequilibrium effect and the averaged effect. The nonequilibrium generation of quantum correlations can be significantly enhanced by large chemical potential (temperature) biases.  This effect is neglected using the Lindblad equation. The \textit{``distilled" nonequilibrium effect}, in contrast to the effect due to the average of the two baths, shows up when the intrinsic time scale of the system is not much greater than the relaxation time, which is exactly when the secular approximation breaks down. When the tunneling rate is small, quantum correlations increase monotonically with the biases of the two reservoirs.  A larger tunneling rate may reduce the entanglement of the system at large biases. When the tunneling rate between two sites is large, the Redfield equation returns approximately the same result as the Lindbladian. Quantum correlations manifests that of an equilibrium behaviors with the average of two baths. Furthermore, at large tunneling regime the quantum entanglement dies at finite temperature bias but does not necessarily vanish with the increase of chemical potential bias. It either suffers a "sudden death" at a finite chemical potential bias or instead asymptotically approaching a non-zero value depending on the value of $\mu_1$. Near the boundary of the extremes, i.e. when the tunneling strength and the site-environment coupling are comparable, we notice the resurrection of the entanglement after its previous drop to zero. 

The analysis in this paper can be generalized to study heat transport or quantum correlations in other systems such as n-level system and spin chains. For a more complicated systems such as spin chains or N-fermionic systems which are more relevant for the development of quantum information processors, the bipartite correlations analyzed in this paper is still applicable by dividing these systems into two non-overlapping subsystems or studying the correlation between two particular sites. In those cases, the transition from small tunneling model to large tunneling model discussed in this study can be calculated numerically, and we still expect similar distinguishable features to appear in the entanglement or other quantum correlations due to the different physics explained in the study. The nonequilibrium enhancement of coherence and entanglement has also been shown to appear in quantum systems such as the qubit system \cite{wuwei1, lgi, spinchain} and three-level systems \cite{cui} though we argue that they are manifestations of the averaged equilibrium effect, and possible experiments can be done to test the nonequilibrium enhancement \cite{59,60, cui}. The results in this study may be useful for designing quantum information devices operating in noisy environment.

\bigskip

\begin{acknowledgments}
X.W and J.W thank the supports from grant no. NSF-PHY 76066 and NSF-CHE-1808474.  X.W. wants to thank Dr. Zhedong Zhang, Kun Zhang and Dr. Wei Wu for helpful discussions, and thank Prof. Dimitri Averin for the questions raised during the presentation of part of the work.
\end{acknowledgments}

\appendix

\section{Quantum correlations for fermionic system in equilibrium environments}\label{equilibrium}
For the system in equilibrium environments, the master equation can be solved exactly. The quantum correlations in equilibrium condition is useful in studying the nonequilibrium case, as part of the correlation generation in the latter case are from the same source as the equilibrium scenario which we want to differentiate. We set the two reservoirs at the same temperature and the same chemical potential. At the long time limit, the system will relax to reach the same equilibrium with the reservoirs without regard to the initial condition. 

\subsection{Bosonic reservoirs} 
\begin{figure}[ht]
\centering
\subfloat[QMI/Discord/CC Vs T]{{\includegraphics[width=.44\textwidth]{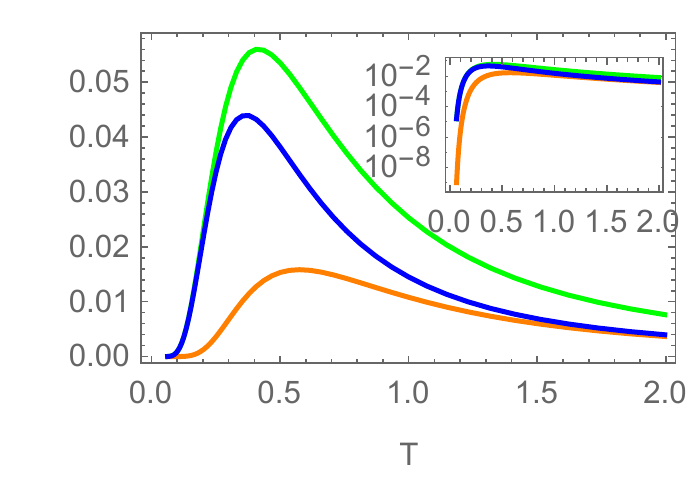}}}
\quad
\subfloat[Concurrence/Discord Vs T]{{\includegraphics[width=.44\textwidth]{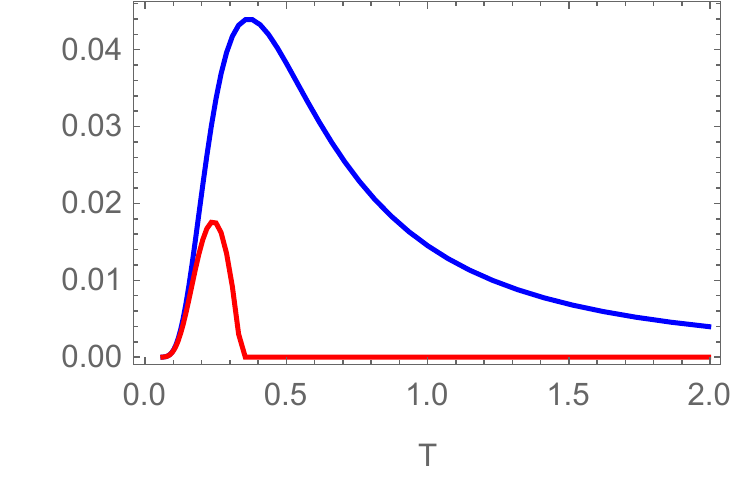}}}
\caption{(Color online) Equilibrium bosonic reservoirs. (a) QMI (green, upper), quantum discord (blue, middle) and CC (orange, lower) Vs temperature (logarithmic scale in the subgraph). (b) Concurrence (red, lower) and quantum discord (blue, upper) and blue Vs T. The parameters are set to $\Delta=0.3$, $\Gamma_1=\Gamma_2=0.05$, $\omega_1=\omega_2=1$.}
\label{fig1}\end{figure}

For bosonic reservoirs, we focus on when the chemical potentials of the reservoirs are zero and the two fermion sites are identical, i.e. $\omega_1=\omega_2$. We set the temperatures of the two reservoirs to be equal, i.e. $T_1=T_2$ and solve for the equilibrium solution of the Redfield equation. The coherence terms of the reduced density matrix for the system vanish and the population in energy basis is given as follows,
\beq \begin{split} \rho_{11}=\frac{(1+ n_1)(1+ n_2)}{(1+2 n_1)(1+2 n_2)}\\ \rho_{22}=\frac{( n_1)(1+ n_2)}{(1+2 n_1)(1+2 n_2)}\\ \rho_{33}=\frac{(1 +n_1)( n_2)}{(1+2 n_1)(1+2 n_2)} \\ \rho_{44}=\frac{(n_1)(n_2)}{(1+2 n_1)(1+2 n_2)}, \end{split} \eeq
where $n_i=n(\omega_i^\prime, T)$ is the population density and $(1+2 n_1)(1+2 n_2)$ on the denominator serves as the normalization factor. The population density above can be simplified to a more recognizable form,
\beq \begin{split} &\rho_{11}=\frac{e^{\hbar \beta (\omega'_1+\omega'_2)}}{Z}, \quad \rho_{22}=\frac{e^{\hbar \beta \omega'_2}}{Z},\\ &\rho_{33}=\frac{e^{\hbar \beta \omega'_1}}{Z}, \qquad \ \rho_{44}=\frac{1}{Z}, \end{split} \label{bosoneq}\eeq
where $Z$ is a normalization factor. The density matrix returns to the result of classical statistics where $\rho_i \propto e^{-\beta E_i}$ in the energy basis, which is expected. There is no coherence term as the result of the decoherence, and the non-vanishing entanglement of the system (see Fig \ref{fig1}) comes from the non-locality of the eigenbasis.

Analysis of the solution shows that with the increase of temperature, all of the quantum correlations we considered, concurrence, discord, classical correlation and mutual information exhibit non-monotonic behavior (see Fig \ref{fig1}). This is easy to understand. At very low temperature, the system which is in equilibrium with the environment, is in its ground state. From previous results \ref{basis}, the ground state of the system $|0\rg \otimes|0\rg$ is localized, which represents no fermion in either site. We expect no quantum correlation in the system of any sort. As the temperature increases, the non-localized excited state becomes more and more dominant. The correlations increase accordingly. In a mixed state, the division between quantum and classical correlation is obscure, and QD and CC has qualitatively similar trend of increasing or decreasing, as shown in Fig \ref{fig1}(a). In the high temperature limit however, all quantum correlations are washed out by the environment and both mutual information and discord decay exponentially.

\begin{figure*}[ht]
\centering
\subfloat[Concurrence/Discord Vs T]{{\includegraphics[width=5.2cm]{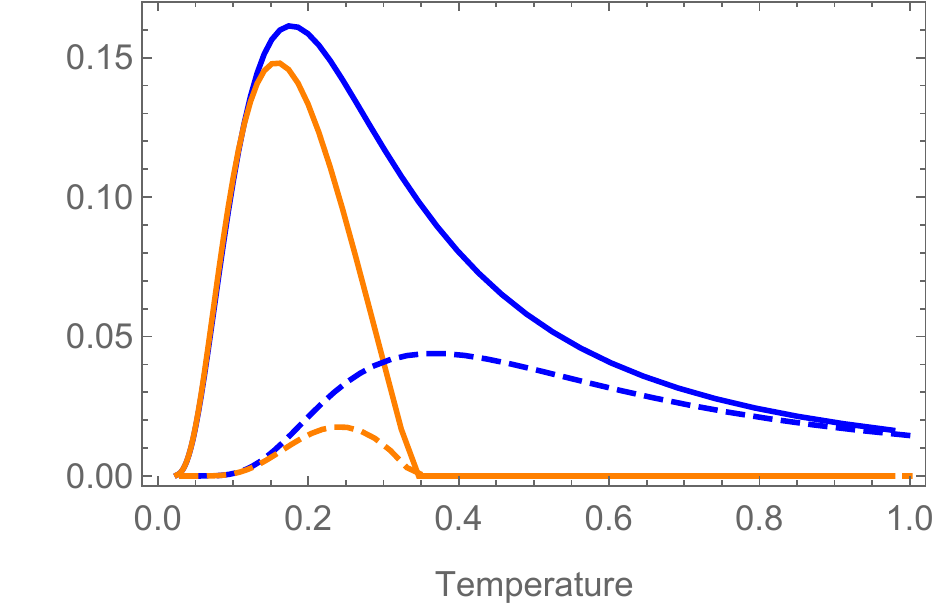}}}
\quad
\subfloat[QD Vs T]{{\includegraphics[width=5.2cm]{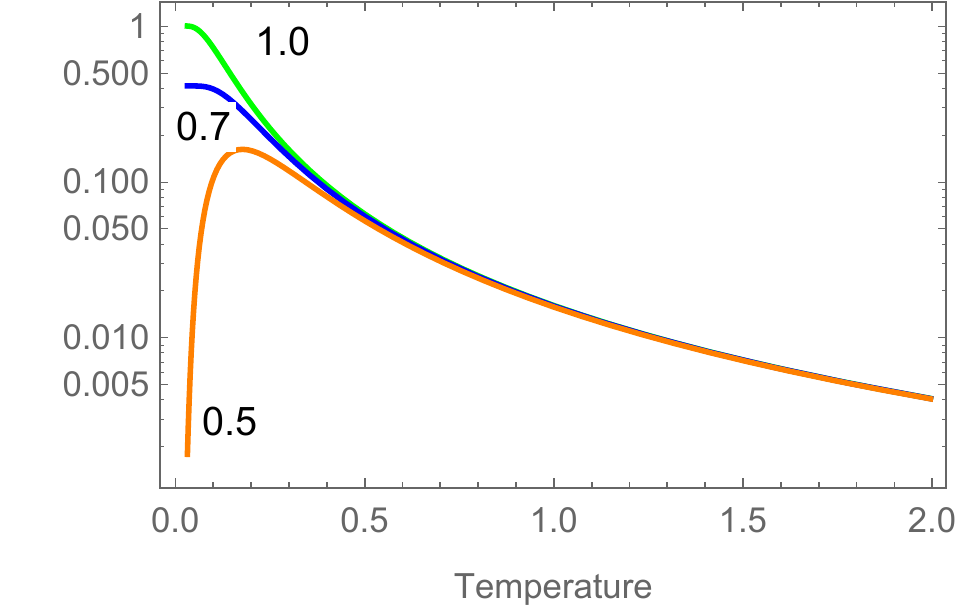}}}
\quad
\subfloat[Concurrence Vs T]{{\includegraphics[width=5.2cm]{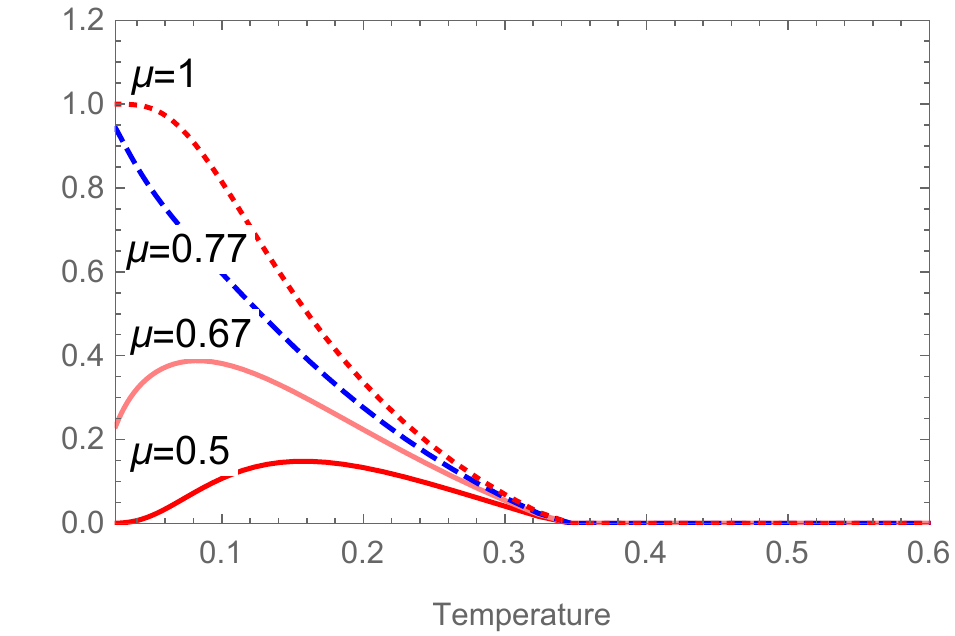}}}
\quad
\subfloat[Concurrenc Vs $\mu$ in log scale]{{\includegraphics[width=5.2cm]{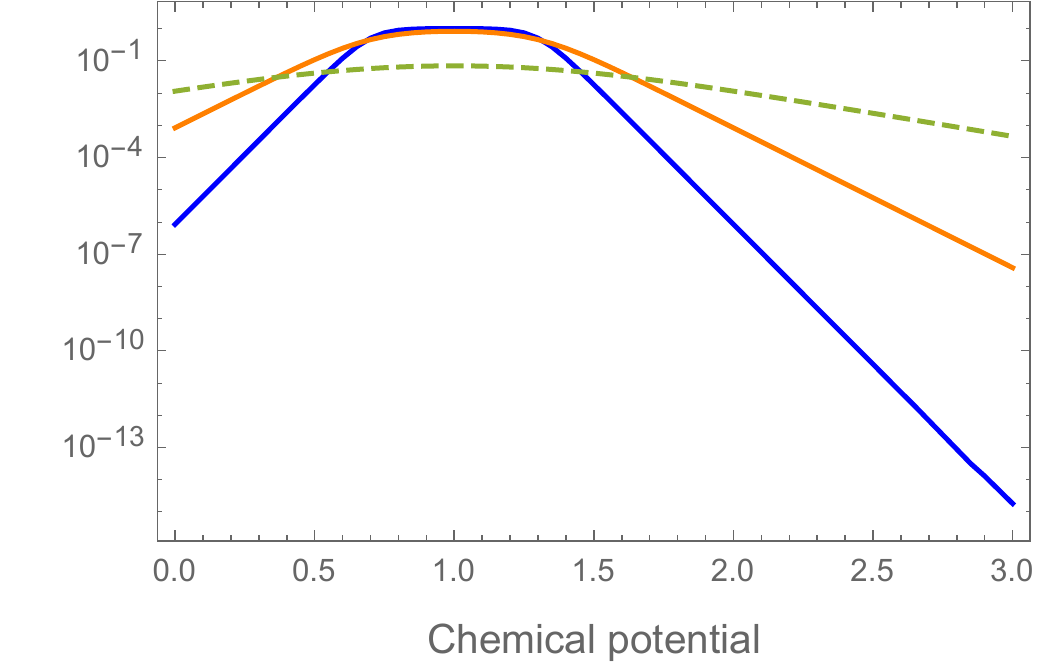}}}
\quad
\subfloat[CC/Discord Vs $\mu$]{{\includegraphics[width=5.2cm]{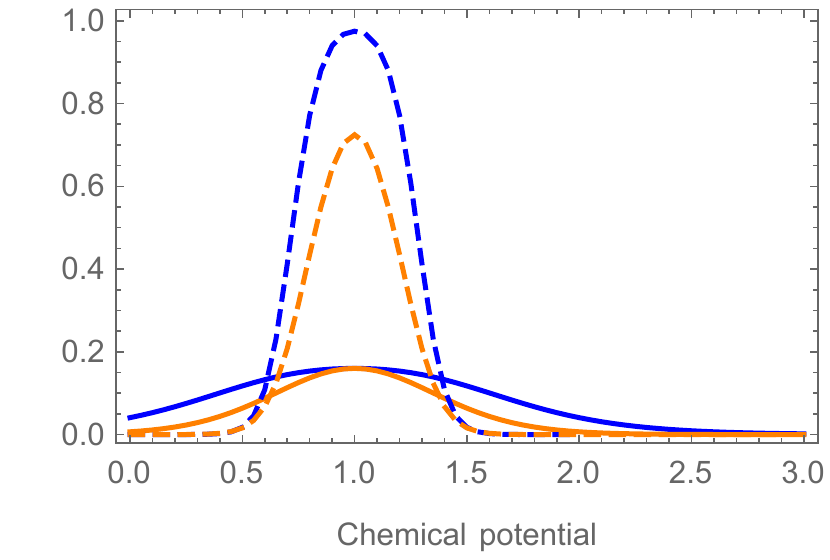}}}
\quad
\subfloat[Discord/Concurrence Vs $\mu$]{{\includegraphics[width=5.3cm]{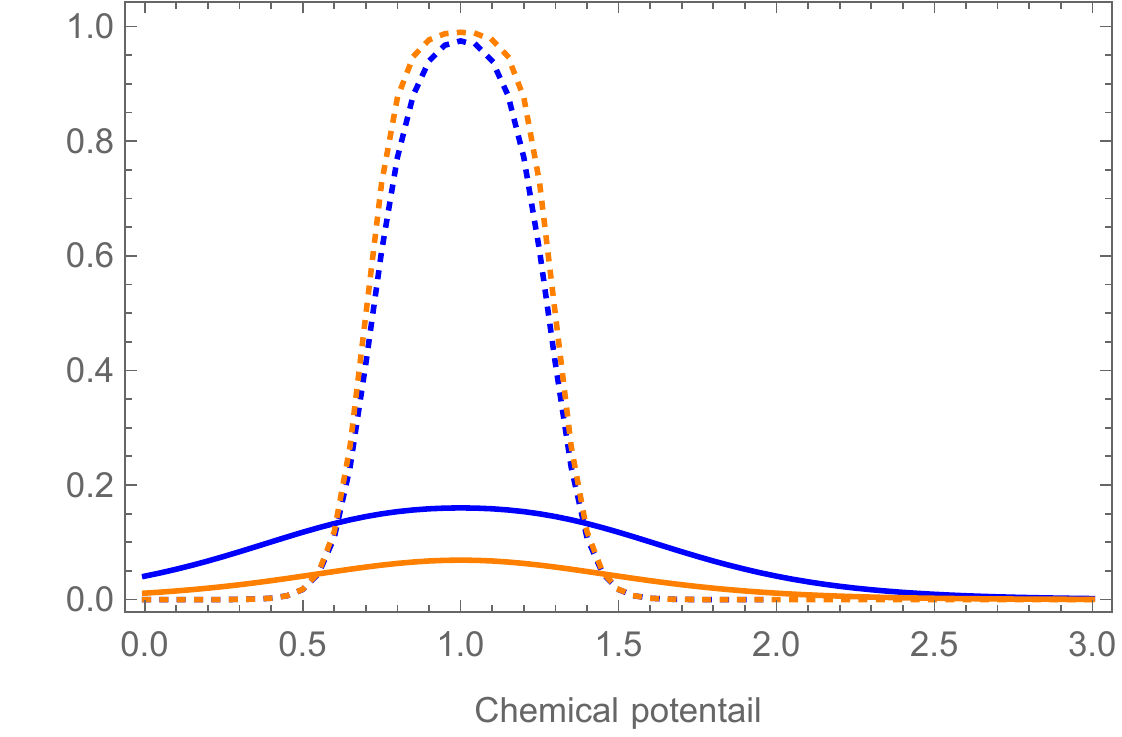}}}
\caption{(Color online) Equilibrium fermionic reservoirs. (a) The  QD (blue) and concurrence (orange) and Vs T with $\mu_1=\mu_2=0.5$ (solid) and $\mu_1=\mu_2=0$ (dashed). (b) Loss of non-monotonicity of QD. (c) Loss of non-monotonicity of concurrence. The turning point is $\mu=0.7$. (d) Concurrence with chemical potential at T= 0.05(blue), 0.1(orange) and 0.3(dashed). (e) CC (orange) and QD (blue) Vs $\mu$ at T=0.1 (dashed) and T=0.3 (solid). (f) QD (blue) and concurrence (orange) Vs $\mu$ at T=0.3 (solid) and T=0.1 (dotted).}
    \label{fig2}
\end{figure*}

On the other hand, the entanglement witnessed a sudden disappearance of it at finite temperature as shown in Fig \ref{fig1}(b). Many abrupt changes of entanglement in the open systems has been reported by many previous studies \cite{YuEberly2,review1, YuEberly, transition}. The disappearance of entanglement is due to the increase of the disorder of the system when temperature is high. Though the entanglement vanishes, the system is still correlated non-classically. This means any local measurement on the subsystem will still perturb the density matrix of the whole system \cite{zurek}.

The concurrence for this given setup $\mathcal{E(\rho)}=2\ \text{Max}(0,|\rho_{23}|-\sqrt{\rho_{11}\rho_{44}})$ can be solved easily. The off-diagonal term $\rho_{23}$ in the local basis representation is directly related to the population terms in the energy basis in the following way $\rho_{23}=(\rho_{33}'-\rho_{22}')/2$ where $\rho_{22}'$ and $\rho_{33}'$ are in the energy representation given in \ref{bosoneq}. The concurrence can be simplified to the following form
\beq \mathcal{E(\rho)}=\frac{e^{\beta \omega}}{Z} \text{Max}(0,\ e^{\beta \Delta}-e^{-\beta \Delta}-2),\eeq
which increases monotonically with the site coupling strength $\Delta$, and it vanishes when the coupling term between the two fermions is weak, i.e. $\Delta<T\ln(1+\sqrt{2})$, or equivalently, when the temperature exceeds the interaction strength \beq T>\Delta/\ln(1+\sqrt{2}), \label{criticalT}\eeq then the system is disentangled. Quantum discord is more robust with respect to the noise from the environment, which suggests that though the system is disentangled when the environment has temperatures slightly above the threshold temperature, the system is still not classical with non-zero quantum correlation. The measurements still disturb the behavior of the system as the quantum.

\subsection{Fermionic reservoirs}

For fermionic environment, we set the two reservoirs at the same temperature and the same chemical potential. We can solve for the equilibrium state of the system, and the population density is given as follows,
\beq \begin{split} &\rho_{11}=(1- n_1)(1- n_2)\\
&\rho_{22}=n_1(1- n_2)\\
&\rho_{33}=(1-n_1) n_2 \\
&\rho_{44}=n_1 n_2, \end{split} \eeq
where $n_i=n(\omega_i^\prime, T)$ is the fermionic occupation number. It is easy to check that thees diagonal density matrix elements also satisfy the classical grand canonical ensemble distribution  $\rho_i \propto e^{-\beta E_i-\mu_i}$.

When the chemical potential is small, the system is almost identical to the system in the bosonic reservoir in the low temperature region as $n_{boson}(T,\omega)=\frac{1}{e^{\beta \omega}-1}\approx n_{fermion}(T, \omega, \mu=0)$ when $\omega/T\gg 1$ (see Fig \ref{fig2}(a)). As we increase the chemical potential of the two baths, the peak of the quantum correlations gradually move to the left until the non-monotonicity completely disappears as is shown in Fig \ref{fig2}(b,c). Intuitively, this can be understood as follows. When the chemical potential increases, the system can capture a fermion more easily from the reservoirs. Due to the tunneling between the two fermion sites, the system gets entangled and correlated. When the chemical potential is close to the energy cost of creating a fermion on the sites, it reaches resonance and quantum correlations are maximized. At the resonant point, thermal fluctuations only play a negative role to the quantum correlation since they wash out the resonance.

The turning point in chemical potential can be calculated by requiring,
\beq \lim_{T\rightarrow 0}\frac{d\ \mathcal{E}(\rho(T,\mu))}{d\ T} < 0
\eeq
and solve for $\mu$. The above inequality gives the turning point, \beq \mu*=\omega-\Delta, \eeq beyond which entanglement only decreases with the temperature. This chemical potential can be calculated exactly for the entanglement, and we can check numerically that the same turning point applies to quantum discord, QMI and CC as well. At this value, the system is in resonance with the lowest non-local eigenstate of the system. And if there is no thermal effect, i.e. T$=0$, the system will in the maximal entangled state where $\mathcal{E}(\rho)=\mathcal{Q}(\rho)=1$ (see Fig \ref{fig2}(d)). As the reservoir gets hot, the degrees of freedom from the environment washes out the coherence between the two fermions. 

Quantum correlations decay exponentially as $\mu$ moves away from the resonant range $[\omega'_1,\omega'_2]$. As an example we plot the concurrence in Fig \ref{fig2}(c)). The exponential behavior of correlations in the low temperature can be understood intuitively as follows. In equilibrium, the system and the reservoirs as a whole share the same chemical potential and temperature. When $T \ll 1$, the expectation value of the particle number on the two non-local energy levels are
\begin{align} n_i(\mu,\beta)=\frac{1}{e^{\beta(\omega_i^\prime-\mu)}+1} \approx e^{\beta(\mu-\omega_i^\prime)\theta(\omega_i^\prime-\mu)},
\end{align}
where $\theta$ is a step function. When $\mu < \omega_1^\prime$, the occupation number $n_2(\mu,\beta) \ll  n_1(\mu,\beta)$ is of higher order of $1/\beta$, and thus is negligible. The occupation number $n_1(\mu,\beta) \approx e^{\beta(\omega_i^\prime-\mu)}$ exponentially increases. When $\omega_1^\prime<\mu<\omega_2^\prime$, the chemical potential is enough to offer one fermion of energy $\omega'_1$ but not enough for $\omega'_2$ fermion. Thus, at $T\ll 1$,  $n_1(\mu,\beta)=1$ and $n_2(\mu,\beta)=0$. The system is essentially in the Bell state which is maximally entangled. When $\mu>\omega_2^\prime$, the system will move to the nearly fully occupied state $|1\rg \otimes |1\rg$, which is localized. $1-n_1(\mu,\beta) \ll 1-n_2(\mu,\beta)=1$, the number of vacancy on the state of energy $\omega_2^\prime$ is $1-\omega_2^\prime=\dfrac{e^{\beta(\omega_2^\prime}}{e^{\beta(\omega_2^\prime-\mu)}+1} \approx e^{\beta(\omega_2^\prime-\mu)}$. When the temperature is low, the two sharp turning points correspond to $\mu_1=\mu_2=\omega_1^\prime$ and $\mu_1=\mu_2=\omega_2^\prime$. As temperature increases, the sharp turnings are smoothed out, and the decay behavior is no longer exponential.

Quantum discord, mutual information and classical correlation all decay exponentially with the increase of temperature. In contrast, concurrence disappears at a finite temperature. It can be shown that the concurrence is
\beq \mathcal{E(\rho)}=\text{Max}(0,\frac{e^{\beta(\omega_1'-\mu)}-e^{\beta(\omega_2'-\mu)}-2e^{\frac{1}{2}\beta(\omega_1'+\omega_2')- \beta\mu}}{(e^{\beta(\omega_1'-\mu)}+1){(e^{\beta(\omega_2'-\mu)}+1)}}), \label{conf} \eeq
and it vanishes at the same threshold temperature \beq T=\Delta/\ln(1+\sqrt{2}) \eeq as in the bosonic reservoir case, which is independent of the chemical potentials.

In the majority of the parameter regimes, the discord is larger than concurrence, however, it is not always true as is shown in Fig \ref{fig2}(f). In the low temperature region, when $\mu \approx 1$, entanglement can be larger than quantum discord. This directly suggests that the discord can not be simply understood as entanglement plus some other non-classical correlations as were pointed out in many studies, e.g. \cite{luo, ali, review2}. Instead,  the discord should be treated as an independent measure for the correlations due to the non-commutative nature of quantum mechanics. 

Comparing with tuning the temperature of the reservoirs to generate quantum correlation in the system, tuning the chemical potential can more directly influence the particle occupation on the two non-local states, generating the correlations with higher and narrower peaks, see Fig \ref{fig2}(e,f). Thermal excitation generates the correlation by perturbing the ground state, and distribute the particle more equally on all four states. Chemical potential generate the correlations through matching the energy of the system with the reservoirs. The maxima of the discord and concurrence all appear at the resonant point when $\mu=(\omega_1'+\omega_2')/2$ at finite temperature.

\begin{widetext}
\bigskip

\section{Density matrix in local basis}\label{lb}

The results we show in this paper are all in Schr\"odinger picture, in Schr\''odinger representation the master equation reads
\begin{equation} \dot \rho_S(t)=\dfrac{i}{\hbar}[\rho_S,H_S]-\dfrac{1}{\hbar^2}e^{-iH_St} Tr_R \int_0^t ds [\tilde H_{int}(t),[\tilde H_{int}(s),\til \rho_S(t)\otimes \rho_R(0)] e^{iH_St}\,.  \label{me}\end{equation}
In the energy eigenbasis, QME takes the form
\begin{equation} \dot \rho_{mn}=\frac{i}{\hbar} (E_n-E_m)\rho_{mn}-\frac{1}{\hbar^2}e^{i(E_n-E_m)t} \cdot \langle m| Tr_R \int_0^t ds \left[\tilde H_{int}(t),[\tilde H_{int}(s),\til \rho_S(t)\otimes \rho_R(0)]\right] |n\rangle\,. \label{mee}\end{equation}

The quantum master equation was solved in the energy eigenbasis, which is a non-local basis. The correlations we studied in the paper are between two localized sites and we perform basis transformation \ref{basis} to the local sites. With the above definition of transformation, the density matrix in the energy eigenbasis $\rho$ can be transformed to the local basis, denoted by $\rho-{local}$ by a unitary transformation given as follows,
\bea  \rho_{local}=U\rho U^\dg =\begin{pmatrix}\rho_{11} & 0 & 0 & 0 \\ 0& c^2 \rho_{22}-cs(\rho_{23}+\rho_{32})+s^2 \rho_{33} & -cs \rho_{22}+s^2 \rho_{23}-c^2\rho_{32}+cs \rho_{33} & 0 \\ 0& -cs \rho_{22}-c^2 \rho_{23}+s^2\rho_{32}+cs \rho_{33} & s^2 \rho_{22}+cs(\rho_{23}+\rho_{32})+c^2 \rho_{33} &0\\0&0&0&\rho_{44} \end{pmatrix}  \eea
where $c=\cos(\theta/2)$ and $s=\sin(\theta/2)$ are transformation angles defined in \ref{newH}.
When the two sites are identical, i.e. $\omega_1=\omega_2$, the above transformation takes a simpler form,
\bea  \rho_{local}=U\rho U^\dg =\begin{pmatrix}\rho_{11} & 0 & 0 & 0 \\ 0& \frac{1}{2}(\rho_{22}+\rho_{33})-\Re (\rho_{23}) & -\frac{1}{2}(\rho_{22}-\rho_{33})+\Im(\rho_{23}) & 0 \\ 0& -\frac{1}{2}(\rho_{22}-\rho_{33})-\Im (\rho_{23}) & \frac{1}{2}(\rho_{22}+\rho_{33}) +\Re(\rho_{23}) &0\\0&0&0&\rho_{44} \end{pmatrix}.  \label{local} \eea

\section{Matrix elements of the master equation}\label{elements}
The operator form of master equation \ref{mst} can be written in terms of its energy eigen basis, and this gives a matrix equation. Here, we give the matrix elements of the differential equations. With the basis defined in \ref{basis}, the master equation \ref{mst} is equivalent to the following form,
\beq \frac{d}{dt}\rho_{ij} = \sum_{lk} M_{ij}^{lk}\rho_{lk}.  \eeq
In the above energy eigenbasis, the nonzero matrix elements $M_{ij}^{lk}$ for bosonic reservoirs is given as follows,
\begin{align}
M_{11}^{11}&=-2(\Gamma_1 (s^2 n_1^{T_1}+c^2 n_1^{T_2})+\Gamma_2 (c^2 n_2^{T_1}+s^2 n_2^{T_2})),\\
M_{11}^{22}&=2\Gamma_1 (s^2 n_1^{T_1}+c^2 n_1^{T_2})+2\Gamma_1,\\
M_{11}^{33}&=2\Gamma_2 (c^2 n_2^{T_1}+s^2 n_2^{T_2})+2\Gamma_2,\\
M_{11}^{23}&=M_{11}^{32}= s c\ (\Gamma_1( n_1^{T_1}- n_1^{T_2})+ \Gamma_2 (n_2^{T_1}- n_2^{T_2})),\\
M_{22}^{11}&=M_{11}^{22}-2\Gamma_1, \\
M_{22}^{22}&=M_{11}^{11}-2\Gamma_1,\\
M_{22}^{44}&=M_{11}^{33}, \\
M_{22}^{23}&=M_{22}^{32}=-M_{11}^{23}+2\Gamma_1(n_1^{T_1}- n_1^{T_2}), \\
M_{33}^{11}&=2\Gamma_2 (c^2 n_2^{T_1}+s^2 n_2^{T_2}),\\
M_{33}^{33}&=M_{11}^{11}-2\Gamma_2,\\
M_{33}^{44}&=M_{11}^{22},\\
M_{33}^{23}&=M_{33}^{32}=-M_{22}^{23},\\
M_{44}^{22}&=M_{33}^{11},\\
M_{44}^{33}&=M_{22}^{11},\\
M_{44}^{44}&=-M_{11}^{22}-M_{11}^{33},\\
M_{44}^{23}&=M_{44}^{32}=-M_{11}^{23}, \\
M_{23}^{11}&=M_{11}^{23}, \\
M_{23}^{22}&=-M_{22}^{23}, \\
M_{23}^{33}&=M_{22}^{23}, \\
M_{23}^{44}&=-M_{11}^{23}, \\
M_{23}^{32}&=M_{11}^{11}-\Gamma_1-\Gamma_2+i(\omega'_2-\omega'_1). 
\end{align}
And for all $l,k$, $M_{32}^{lk}=(M_{23}^{lk})^*.$ The rest of the matrix elements are zero. The other off-diagonal terms are uncoupled to the population density matrix elements, and thus disappear in the steady state due to the decoherence.

For fermionic reservoirs,  the nonzero matrix elements $M_{ij}^{lk}$ are
\begin{align}
M_{11}^{11}&=-2(\Gamma_1 (s^2 n_1^{T_1}+c^2 n_1^{T_2})+\Gamma_2 (c^2 n_2^{T_1}+s^2 n_2^{T_2})),\\
M_{11}^{22}&=-2\Gamma_1 (s^2 n_1^{T_1}+c^2 n_1^{T_2})+2\Gamma_1,\\
M_{11}^{33}&=-2\Gamma_2 (c^2 n_2^{T_1}+s^2 n_2^{T_2})+2\Gamma_2,\\
M_{11}^{23}&=M_{11}^{32}= - s c\ \big(\Gamma_1( n_1^{T_1}- n_1^{T_2})+ \Gamma_2 (n_2^{T_1}- n_2^{T_2})\big),\\
M_{22}^{11}&=-M_{11}^{22}+2\Gamma_1, \\
M_{22}^{22}&=-M_{11}^{22}+M_{11}^{33}-2\Gamma_2,\\
M_{22}^{44}&=M_{11}^{33}, \\
M_{22}^{23}&=M_{22}^{32}=-M_{11}^{23}, \\
M_{33}^{11}&=2\Gamma_2 (c^2 n_2^{T_1}+s^2 n_2^{T_2}),\\
M_{33}^{33}&=M_{11}^{22}-2\Gamma_1-M_{11}^{33},\\
M_{33}^{44}&=M_{11}^{22},\\
M_{33}^{23}&=M_{33}^{32}=-M_{11}^{23},\\
M_{44}^{22}&=M_{33}^{11},\\
M_{44}^{33}&=M_{22}^{11},\\
M_{44}^{44}&=-M_{11}^{22}-M_{11}^{33},\\
M_{44}^{23}&=M_{44}^{32}=M_{11}^{23}, \\
M_{23}^{11}&=M_{23}^{22}=M_{23}^{33}=M_{23}^{44}=-M_{11}^{23}, \\
M_{23}^{32}&=-\Gamma_1-\Gamma_2+i(\omega'_2-\omega'_1).
\end{align}
$M_{32}^{lk}=(M_{23}^{lk})^*$ for all $l,k$ and the rest of the matrix elements are zero.

\section{Solution of Nonequilibrium Master equations}
In the section \textit{Quantum Correlations for Fermionic System in Nonequilibrium Environments}, we gave the solution for the master equations up to the leading order for the convenience of analysis. Here, we present the exact solutions which are useful for checking our numerical results and when the order expansion is not applicable.
\subsection{Solution of bosonic reservoirs} \label{sb}
In the energy eigenbasis, density matrix of the nonequilibrium steady state for bosonic environment is given as follows,
\begin{align*}
\rho_{11}=\frac{1}{N}
\Big\{\Gamma^2& \Big[16 + 3n_{2m}^2 + 24 n_{2p} + n_{1p}^3 (2 + n_{2p}) + 2 n_{1p}^2 (2 + n_{2p}) (3 + n_{2p}) + n_{1m}^2 (3 + n_{1p} + n_{2p})-\\
      & n_{1m} n_{2m} (10 + n_{1p}^2 + 2 n_{1p} (3 + n_{2p}) + n_{2p} (6 + n_{2p}))+  n_{2p} (n_{2m}^2 + 2 n_{2p} (6 + n_{2p})) + \\
      & n_{1p}\Big(n_{2m}^2 + (2 + n_{2p})^2 (6 + n_{2p})\Big)\Big] + (2 + n_{1p}) (2 + n_{2p}) \omega_{12}^{\prime 2}\Big\} \\
\rho_{22}=\frac{1}{N}
\Big\{g^2& \Big[n_{1p}^3 (2 + n_{2p}) + 2 n_{1p}^2 (2 + n_{2p})^2 - n_{2m}^2 (3 + n_{2p}) + n_{1m}^2 (1 + n_{1p} + n_{2p}) + n_{1p} (-n_{2m}^2 + (2 + n_{2p})^3) - \\
      & n_{1m} n_{2m} \Big(2 + n_{1p}^2 + 2 n_{1p} (2 + n_{2p}) + n_{2p} (4 + n_{2p})\Big)\Big]+n_{1p} (2 + n_{2p}) \omega_{12}^{\prime 2}\Big\} \\
\rho_{33}=\frac{1}{N}
\Big\{\Gamma^2& \Big[ n_{2m}^2 (1 + n_{1p} + n_{2p}) - (2 + n_{1p}) n_{2p} (2 + n_{1p} + n_{2p})^2 + n_{1m}^2 (3 + n_{1p} + n_{2p}) + \\
	& n_{1m} n_{2m} \Big(2 + n_{1p}^2 + 2 n_{1p}(2 + n_{2p}) + n_{2p} (4 + n_{2p})\Big)\Big]+ (2 + n_{1p}) n_{2p} \omega_{12}^{\prime 2}\Big\} \\
\rho_{44}=\frac{1}{N}
 \Big\{\Gamma^2& \Big[ n_{1p }n_{2p} (2 + n_{1p} + n_{2p})^2 - n_{1m}^2 (1 + n_{1p} + n_{2p}) - n_{2m}^2 (1 + n_{1p}+ n_{2p}) - \\
	& n_{1m} n_{2m}\Big(2 + n_{1p}^2 + 2 n_{1p} (1 + n_{2p}) + n_{2p} (2 + n_{2p})\Big)\Big] + n_{1p} n_{2p }\omega_{12}^{\prime 2}\Big\} \\
\rho_{23}=\frac{1}{N}
\Big\{2&\Gamma^2 (2 + n_{1p} + n_{2p}) (n_{1m} + n_{2m} + n_{1p} n_{2m} + n_{1m} n_{2p}) -  \\
&-2i\Gamma \big(n_{1m} + n_{2m} + n_{1p} n_{2m} + n_{1m} n_{2p}\big) \omega'_{12} \Big\}
\end{align*}
where normalization factor is $N=4 \Gamma^2 (2 + n_{1p} + n_{2p})^2 (1 + n_{1p} - n_{1m} n_{2m} + n_{2p} + n_{1p} n_{2p}) - 4 (1 + n_{1p}) (1 + n_{2p}) \omega_{12}^{'2}$, $n_{ip}=n(\omega'_i,T_1)+n(\omega'_i,T_2)$, $n_{im}=n(\omega'_i,T_1)-n(\omega'_i,T_2)$ with i=1,2 and $\omega'_{12}=\omega_1'-\omega'_2$

\subsection{Solution of fermionic reservoirs} \label{sf}
In the fermionic reservoirs, the density matrix of the nonequilibrium steady state in the energy eigenbasis is laid down as the follows,
\begin{align*}
\rho_{11}=\frac{1}{M} & \{ \Gamma^2  (4 (2 -  n_{1p}) (2 - n_{2p})-(n_{1m} + n_{2m})^2) + (2 - n_{1p}) (2 -n_{2p}) \omega_{12}^{\prime 2}\} \\
\rho_{22}=\frac{1}{M} &\{ \Gamma^2  ((n_{1m} + n_{2m})^2 + 4  n_{1p} (2 - n_{2p})) + n_{1p} (2 - n_{2p}) \omega_{12}^{\prime 2}\}\\
\rho_{33}=\frac{1}{M} &\{ \Gamma^2  ((n_{1m} + n_{2m})^2 + 4 (2 -  n_{1p}) n_{2p}) + (2 -  n_{1p}) n_{2p} \omega_{12}^{\prime 2}\}\\
\rho_{44}=\frac{1}{M} &\{ \Gamma^2  (4  n_{1p} n_{2p}-(n_{1m}+ n_{2m})^2 ) +  n_{1p} n_{2p} \omega_{12}^{\prime 2}\}\\
\rho_{23}=\frac{1}{M} &\{ 4\Gamma^2  (n_{1m} + n_{2m})-2i\Gamma (n_{1m} + n_{2m}) \omega_{12}^\prime \},
\end{align*}
where normalization factor $M=4 (4\Gamma^2 + {\omega^\prime_{12}}^2)$, $n_{ip}=n(\omega'_i,T_1,\mu_1)+n(\omega'_i,T_2,\mu_2)$, $n_{im}=n(\omega'_i,T_1,\mu_1)-n(\omega'_i,T_2,\mu_2)$ with i=1,2 and $\omega'_{12}=\omega_1'-\omega'_2$.

\section{Energy current}\label{current}

\begin{figure}[h]
\centering
\includegraphics[width=0.45 \textwidth]{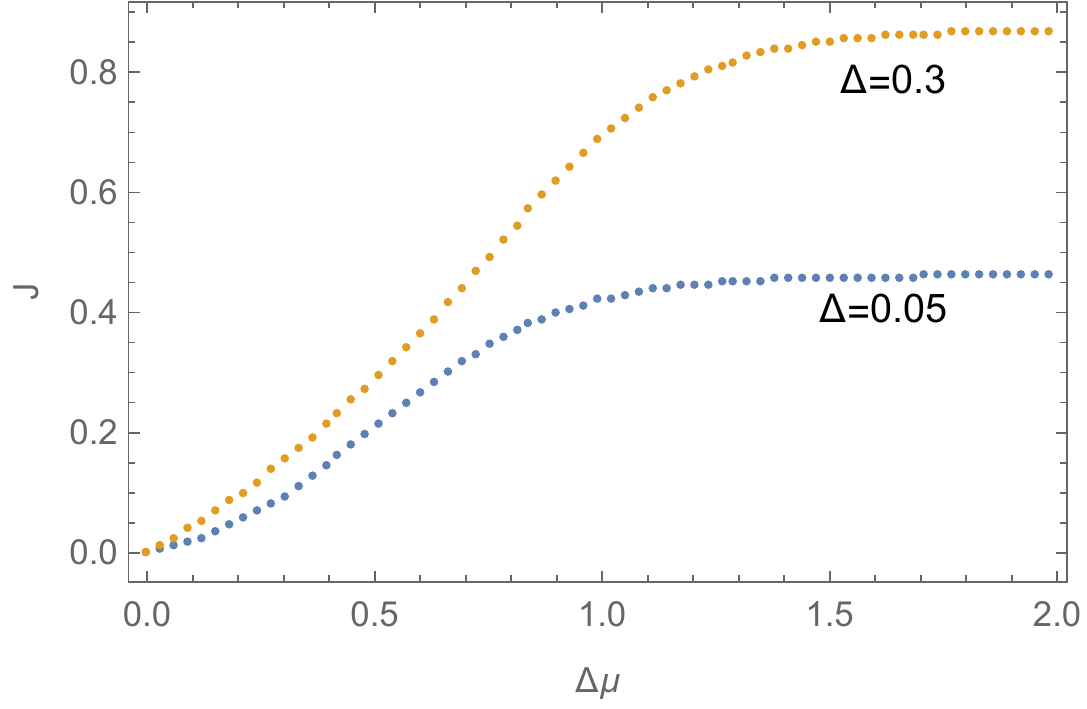}
\caption{(Color online) Energy current at $\Delta=0.05$ (lower, blue) and 0.3 (upper, orange).  The parameters are $T_1=T_2=0.2$, $\mu_1=0.5$, $\mu_2=\mu_1+\Delta T$, $\Gamma_{1,2}=0.05$ and $\omega'_{1,2}=1\pm\Delta$.}
\label{crnt}
\end{figure}

The appearance of constant energy current is one of the features of the nonequilibrium system. The energy flow in our case is from one reservoir of higher potential or temperature to the system and then to the reservoir with lower potential or temperature. The definition of energy current is the following. The master equation in \ref{mst} can be decomposed in the following form,
\beq \dot \rho_S(t)=i[\rho_S,H_S]-D_0[\rho]-D_s[\rho]=i[\rho_S,H_S]+\sum_{i=1,2}D_{i}[\rho]  \eeq
where $D_i[\rho]$ is the dissipator in contact with the $i^{th}$ reservoir. The steady state energy current is defined as \cite{thermal, wuwei1,cui}
\beq J_i=\text{Tr}\{D_i [\rho_{ss}] \calh_S\},  \eeq
where the index $i$ means that the current flows from the $i^{th}$ reservoir to the system. In the steady state, the incoming energy current and the outgoing energy current have to balance out, i.e. $J_1=-J_2$. Therefore, without loss of generality, we will only calculate $J_1$. 

The explicit expression for $J_1$ is given as below. For bosonic reservoir,
\begin{align}
J_1=&-2 \omega'_1 \Gamma_1 s^2 [(1+n_1^{T_1})(\rho_{22}+\rho_{44})-n_1^{T_1}(\rho_{11}+\rho_{33})]-2 \omega'_2 \Gamma_2 c^2 [(1+n_2^{T_1})(\rho_{33}+\rho_{44})-n_2^{T_1}(\rho_{11}+\rho_{22})]- \nonumber \\
&-\omega'_2 \Gamma_1 sc\ [(1+n_1^{T_1})(\rho_{22}+\rho_{33})-n_1^{T_1}(\rho_{22}+\rho_{33})]-\omega'_1 \Gamma_2 sc\ [(1+n_2^{T_1})(\rho_{22}+\rho_{33})-n_2^{T_1}(\rho_{22}+\rho_{33})]. 
\end{align}
For fermionic reservoir, 
\begin{align}
J_1=&-2 \omega'_1 \Gamma_1 s^2 [(1-n_1^{T_1})(\rho_{22}+\rho_{44})-n_1^{T_1}(\rho_{11}+\rho_{33})]-2 \omega'_2 \Gamma_2 c^2 [(1-n_2^{T_1})(\rho_{33}+\rho_{44})-n_2^{T_1}(\rho_{11}+\rho_{22})]- \nonumber \\
&-\omega'_2 \Gamma_1 sc\ (\rho_{22}+\rho_{33})-\omega'_1 \Gamma_2 sc\ (\rho_{22}+\rho_{33}),
\end{align}
where all $\rho_{ij}$'s above are elements of density matrix in energy basis. 

As an example, we plot the current with the change of chemical potential bias in fermionic reservoir case. The current saturates at a lower value when the tunneling between sites is smaller, see Fig\ref{crnt}. As the chemical potential bias increases, the energy current increases as expected. The line with smaller tunneling rate saturates at smaller bias and reaches a smaller asymptotic value. 
\end{widetext}

\bigskip

\end{document}